\documentclass[12pt]{JHEP3}
\usepackage{amssymb,amsfonts}
\usepackage{amsmath}
\usepackage{graphicx}
\usepackage{amsmath,epsfig}
\usepackage{amssymb,amsfonts}
\usepackage{latexsym}
\def\refeq#1{(\ref{#1})}
\def\hri#1#2{\href{http://arxiv.org/abs/#1}{[ArXiv:#1]#2}}
\def\hre#1#2{\href{http://arxiv.org/abs/#1/#2}{[ArXiv:#1/#2]}}

\def\pa{\partial}
\def\dt{\partial}
\def\diag{{\rm diag}}

\def\cG{{\cal G}}
\def\be{\begin{equation}}
\def\ee{\end{equation}}
\def\bea{\begin{eqnarray}}
\def\non{\nonumber}

\def\eea{\end{eqnarray}}
\def\sp{\;\;\;,\;\;\;}
\def\l{\lambda}
\def\im{{\rm Im}}
\def\cN{{\cal N}}
\def\lab{\label}

\def\o{\omega}
\def\le{\left}
\def\ri{\right}

\def\half{\frac12}

\def\cO{{\cal O}}

\def\n{\nu}
\def\td{\tilde}
\def\d{\delta}
\def\6{\partial}
\def\de{\partial}

\def\ls{\ell_s}
\def\a{\alpha}
\def\b{\beta}
\def\d{\delta}
\def\G{\Gamma}
\def\lab{\label}
\def\parl{\parallel}

\def\r{x}

\def\dx{\delta X^\parl}

\def\<{\langle}
\def\>{\rangle}

\def\g{\gamma}
\newcommand{\ud}{\mathrm{d}}

\title{\vskip -1cm Dressed spectral densities for heavy quark diffusion in holographic plasmas}
\author{Elias Kiritsis$^{1,3}$, Liuba Mazzanti$^2$ and Francesco Nitti$^1$\\
~\\
$^1$\href{http://www.apc.univ-paris7.fr}{APC, Universit\'e Paris 7, B\^atiment Condorcet, F-75205, Paris Cedex 13, France,
 (UMR du CNRS 7164).}\\
~\\
$^2$\href{http://www-fp.usc.es/~theory/}{Departamento de F\'{\i}sica de Part\'{\i}culas, Universidade
de Santiago de
Compostela\\and Instituto Galego de F\'{\i}sica de Altas
Enerx\'{\i}as (IGFAE)\\E-15782, Santiago de Compostela, Spain}\\
~\\
 $^3$\href{http://hep.physics.uoc.gr/}{Crete Center for Theoretical Physics,\\Department of Physics, University of Crete
71003 Heraklion, Greece}\\
 }
\preprint{CCTP-2011-29}
\abstract{We analyze the large frequency behavior of the spectral densities that govern the generalized Langevin diffusion process for a heavy quark
in the context of the gauge/gravity duality. The bare Langevin correlators obtained from the trailing string solution have a singular short-distance
behavior. We argue that the proper dressed spectral functions are obtained by subtracting the zero-temperature correlators. The dressed spectral functions have a sufficiently fast fall-off at large
frequency so that the Langevin process is well defined and the dispersion relations are satisfied. We identify the cases in which the
subtraction does not modify the associated low-frequency transport coefficients. These include conformal theories and the non-conformal,
non-confining models. We provide several analytic and numerical examples in conformal and non-conformal holographic backgrounds.}
\vspace{-1cm}
\keywords{\vspace{-1cm} AdS/CFT, Quark-Gluon Plasma,Langevin diffusion, heavy quarks}

\begin{document}

\section{Introduction and results}\label{INTRO}

Heavy-ion collision experiments and related data on the deconfined phase of QCD, \cite{rhic}, have provided a first window for string theory techniques
to meet the real world. The context is strong coupling dynamics near and above the deconfining transition in QCD. String theory via the $AdS$/CFT
correspondence has provided a framework in order to understand strong coupling dynamics in the deconfined phase including the calculation of transport
coefficients. Recent reviews on the progress in this direction are \cite{sonrev,csa,iancurev,gubserrev,ihqcdrev}.

Observables of particular importance are associated to heavy quarks. Heavy quarks may be produced in the Quark-Gluon Plasma (QGP) of the RHIC
fireball and are then traveling to the detectors while moving through the dense QGP. They can be tagged reasonably well and are therefore valuable
probes of the dynamics in the plasma and in particular for the mechanism of energy loss.

A single heavy quark can be modeled in string theory by an open string. Its end-point is representing the heavy quark while the string is trailing
behind as the quark moves. The large mass limit is important in order to neglect the non-trivial flavor dynamics associated with light quarks
(although with improved techniques the light quarks may also eventually be addressed reliably in the holographic context). As quarks are associated
with strings ending on flavor branes, a heavy quark ends on a brane that is stretching in the UV part of the bulk geometry. The motion of such a
string, and the associated force acting on the quark from the thermal medium, have been studied in detail with several complementary methods,
\cite{her,lrw,gub1}. In the simplest setup, the UV endpoint of a fundamental string is forced to move with constant velocity $v$ along a spacial
direction. The equations of motion for the full string are solved and the radial profile of the trailing string is found as it moves in a bulk
black hole background representing the deconfined heat bath. The energy absorbed by the string is calculated and the drag force of the string is
obtained. The picture remains roughly valid, while details change when conformal invariance is broken, \cite{lliu,transport}.

An important improvement in this picture consists of the study of the stochastic nature of this system, in analogy with the dynamics of heavy
particles in a heat bath giving rise to Brownian motion. This involves a diffusive process, that was first considered in a holographic setting in
\cite{tea}, by using the Schwinger-Keldysh formalism adapted to $AdS$/CFT in \cite{sonherzog}.

Subsequently, a study of the (quantum) fluctuations of the trailing string, \cite{gubser,Casal}, provided the information on the momentum broadening of
a heavy quark as it moves in the plasma. The stochastic motion was formulated as a Langevin process, \cite{deboer,sonteaney}, associated with the
correlators of the fluctuations of the string.

Many heavy quarks in experiments are relativistic. Therefore it is necessary to study the associated relativistic Langevin evolution of the trailing
string, a feat accomplished in the ${\cal N}=4$ case in \cite{iancu} and in general non-conformal plasmas in \cite{hoyos,langevin-1}. The same type of
Langevin process was studied in \cite{guijosa} for the case of an accelerating quark in the vacuum (rather than in a deconfined plasma), by analyzing
the fluctuations of a trailing string in $AdS$ with a non-uniformly moving endpoint.

On the experimental front, there have been several results from the RHIC experiments, \cite{ph0}-\cite{ph3}. The experimental signatures are
currently summarized by the $e^{\pm}$ spectra that originate in the semileptonic decays of charmed and bottom hadrons. From these spectra a
modification factor $R^e_{AA}$ and an elliptic flow coefficient $v^e_2$ are extracted. They capture the effects of the medium to the propagation of
the heavy quarks. The data exhibit a substantial elliptic flow, up to $v_2^e\simeq$ 10\%, and a high-$p_T$ suppression down to $R^e_{AA}\simeq 0.25$.
These values are comparable to light hadrons. Radiative energy loss models based on pQCD, \cite{dainese}, do not seem to explain well the experimental
data, \cite{rapp}. Elastic scattering energy loss plus non-perturbative interactions can on the other hand accommodate the data, \cite{rapp}.

In particular, the Langevin approach has been applied to the study of the heavy quark energy loss by several groups, and the related physics is
summarized in the review \cite{rapp}. The Langevin evolution used was relativistic and with symmetric diffusion coefficients. As there was no
microscopic model to provide the proper fluctuation-dissipation relation, the Einstein equations used vary, and in all examples it was assumed that
the equilibrium distribution is the J\"uttner-Boltzmann distribution. Moreover various combinations of friction forces were used, resonance models,
pQCD, $\cal N$=4 $AdS$/CFT and combinations. A further recent analysis was performed in \cite{lan} with similar conclusions. The associated relativistic and
isotropic Langevin systems used have been introduced in the mathematical physics literature rather recently, \cite{rel} (see \cite{rel1} for a
review).

Before proceeding to the issues studied in this paper, we present briefly the results of \cite{langevin-1} in order to put our conclusions here in
context. In \cite{langevin-1} a large class of non-conformal backgrounds captured by Einstein-dilaton gravity with a dilaton potential in 5
dimensions were considered in the context of heavy-quark energy loss. In a series of recent works, such backgrounds were analyzed both qualitatively
and quantitatively and have provided a rich variety of holographic bulk dynamics. In particular, for a selected class of scalar potentials, they mimic
the behavior of large-$N$ Yang Mills, \cite{ihqcd1}-\cite{gkmn3}. This match can be quantitative, \cite{gkmn3}, agreeing very well both at zero and
finite temperature with recent high-precision lattice data, \cite{panero}. However the Langevin analysis was done for generic non-conformal
backgrounds in \cite{langevin-1}.

The main method is to consider a string end-point that is forced to move with velocity $v$. Solving the Nambu-Goto equations of motion, the classical
profile of the trailing string can be found. The string stretches inside the bulk until it becomes completely horizontal at some value of radial
coordinate $r_s$, given by $f(r_s)=v^2$ where $f(r)$ is the blackness function of the background. When the quark is moving slowly, as $v\to 0$, the
point $r_s$ approaches the bulk black hole horizon.

The induced metric on the string world-sheet has the form of a two-dimensional black hole metric with a horizon at $r=r_s$ as first observed in
\cite{Casal}.\footnote{This is a generic effect on strings and D-branes embedded in black hole/black-brane backgrounds. It was first observed in
\cite{cli} where it was used to propose that a different speed of light is relevant for such branes. It is implicit or explicit in many holographic
computations using probe flavor branes, \cite{karch,mateos} and strings \cite{Casal}.} This black hole is an important ingredient of the dynamics of
the system. In particular it is crucial in the calculation of the thermal correlators using the Schwinger-Keldysh formalism, as well as for the
fluctuation-dissipation relation. The world-sheet black hole has an associated Hawking temperature $T_s$ that depends on several parameters: the
background temperature $T$, the zero-temperature bulk scale $\Lambda$ and the quark velocity $v$. It coincides with the temperature $T$ of the heat
bath only in the non-relativistic limit. In the conformal case, one has $T_s= T_{\rm s,conf} = T(1-v^2)^{1\over 4}\leq T$. The numerical analysis of
many examples shows that $T_s\leq T_{\rm s,conf}\leq T$. The equality $T_s=T_{\rm s,conf}$, in the first relation is attained, for arbitrary $v$, in
the high T limit and also in the ultra-relativistic limit, $v\to 1$.

The small fluctuations around the classical string profile satisfy second-order radial equations that are related to the associated thermal
correlators by the holographic prescription. It should be emphasized that such correlators are thermal with temperature $T_s$ and not the temperature
$T$ of the heat bath. Moreover, they satisfy the fluctuation-dissipation relation associated with the emergent temperature $T_s$. The fact that the
string fluctuations see a modified temperature crucially affects the Einstein relation between the diffusion constants.

At the quadratic level of fluctuations, a relativistic Langevin diffusion equation is obtained using the $AdS$/CFT prescription.
The properties of this relativistic Langevin evolution differ substantially from rotationally invariant equations that have been introduced recently
in mathematical physics \cite{rel,rel1,lan}. In particular, here the evolution is not rotationally symmetric, and the Einstein relation is different,
because the fluctuation-dissipation relation is different. This implies that the equilibrium configuration is not the standard rotationally-invariant
J\"uttner-Boltzmann distribution.

The main focus of \cite{langevin-1} was on the local Langevin equation, which arises when looking at the large-time limit of the fluctuations of the
heat bath, i.e. at the small frequency modes. On the other hand, the holographic computation gives access to the full frequency spectrum of the
correlation functions driving the generalized Langevin dynamics. The holographic computation of the full Langevin correlators and the associated
spectral densities for improved Holographic QCD was done mostly numerically in \cite{langevin-1}. Analytic expressions (in terms of the bulk metric
and dilaton profiles) were obtained in the two opposite regimes of small and large frequencies $\omega$ (compared to an appropriate temperature scale)
for general non-conformal backgrounds.

In the large-frequency regime, the spectral densities are obtained via a modified WKB method, similar to the one followed in \cite{Teaney} for {\em
bulk} fluctuations in an $AdS$-Schwarzschild background. The high-frequency behavior is different, depending on the mass of the probe quark. For
finite mass, and for large $\omega$, the spectral densities grow linearly with $\omega$, whereas in the limit when the quark mass becomes infinite
this behavior changes to a cubic power-law. However as we will see later in this work, a part of this regime is unreliable.

Going beyond the zero-frequency limit is necessary when the diffusion process happens on time scales comparable to, or smaller than the
auto-correlation time of the fluctuation propagators. More specifically, since these are thermal correlators at the temperature $T_s$, the large-time
approximation breaks down over time-scales shorter than $T_s^{-1}$. This condition puts a temperature-dependent upper bound on the momenta of the
heavy quark, above which the diffusion process cannot be described by a simple {\em local} Langevin equation with white noise. In this context, it is
useful to have approximate expressions valid for large frequencies (for examples, those we obtain with the WKB method) to model the behavior of the
system in the regime where the local Langevin approximation breaks down and the dynamics becomes non-markovian (due to a non-trivial memory kernel)

The numerical evaluation of the diffusion constants may lead directly to a comparison of the jet-quenching parameters between the holographic QCD
model and data. It was found that $\hat{q}_\perp$ displayed a mild momentum dependence for large quark momenta, which however differs from the one
obtained holographically in the conformal case. As the temperature rises, $\hat{q}_\perp$ increases significantly, approximately as $\sim T^3$.
Interestingly, it was found that for temperatures above $\sim 400~MeV$, the local description of the diffusive process {\em breaks down} for charm
quarks with momenta above $\sim 5-10~GeV$. This is because the process occurs on time scales shorter than $1/T_s$. This implies that in order to
describe heavy charm quark diffusion in the ALICE experiment, one would need the full generalized non-local Langevin equation, and the full
frequency-dependent correlator, rather than just its low-frequency limit captured by $\hat{q}_\perp$. This would constitute an interesting testing
ground for holographic models, where the full correlators can be easily computed. The energies at which the energy loss mechanism described, is not
any more the dominant one and radiation becomes the dominant mechanism were also determined. This was estimated by requiring that $r_s$ remains below
the would-be position of the flavor brane, \cite{Casal,lrwr}. It was shown that these limit do not substantially constrain this framework.

The issues analyzed in this paper arose from the intent to apply the results of \cite{langevin-1} to comparison with experimental data. It was argued
in \cite{langevin-1} that for parts of the phase space of heavy quarks in LHC, the full non-Markov Langevin is relevant, and therefore the full
Langevin correlators are needed. When trying to apply this Langevin evolution to real quarks, two immediate questions arise:

\begin{itemize}

\item What is the proper behavior of the correlators at high-frequency $\omega$, and how is this expected to affect heavy quark diffusion?

\item What is the proper observable motion of heavy quarks in a strongly coupled plasma?

\end{itemize}

A direct derivation of the Langevin evolution in the vacuum of the strongly coupled theory (at $T=0$), involves the zero temperature retarded and
symmetric correlators. As usual the retarded correlator controls the traditional dissipative term while the symmetric correlator the fluctuations of
the Langevin noise.

As shown in earlier works, \cite{Casal,gubser}, the retarded $T=0$ correlator in $AdS$, is $G_R\sim \omega^3$ (in the infinite quark mass limit). The same
is the large frequency behavior in any asymptotically $AdS$ case, \cite{langevin-1}. This short distance behavior provides a Langevin evolution at
$T=0$ with a ``dissipative force'' that contains the third time derivative of the coordinate. Obviously this behavior is unphysical and the short
distance behavior of the Langevin evolution must be redefined so that they is no semiclassical dissipation in the ground state ($AdS$ vacuum).

A related issue concerns the causality of the evolution ingrained into the causality-related dispersion relations (\ref{dispersion}) for the force
correlators. Such dispersion relations are valid only if the spectral densities vanish sufficiently fast at large frequencies.

The above considerations suggest that the proper definition of the Langevin diffusion of the heavy quark in the plasma must be defined so that the
diffusion in the vacuum is trivial. This involves the redefinition (dressing) of heavy quark coordinates together with an appropriate modification of
the path integral measure controlling their fluctuations.

This procedure is developed from first principles in this paper. We find that in the regime in which the fluctuations of the quark coordinates are
perturbative, the retarded correlator $G_R(\omega,T)$ is replaced with $G_R(\omega,T)-G_R(\omega,0)$ which vanishes at $T=0$ and falls off (as $1/\o$)
at large $\omega$. The corrections to this simple result are controlled by ${\gamma \omega \over M_Q}\ll 1$. This condition is necessary because, in
the gravity picture, having $\gamma \o \gg M_Q$ implies a breakdown of the boundary effective description of the string endpoint as a free heavy
quark. Also, in this regime non-linear backreaction effects become important.

Similarly, the symmetric correlator is replaced by $G_s(\omega,T)-G_s(\omega,0)$. the Einstein relation relating $G_{R}$ to $G_s$ gets modified, but
it unchanged in the zero-frequency limit. Therefore, the relation between the friction and noise term in the local limit of the Langevin process is
the same as in \cite{langevin-1}.

A subtraction similar to the one we perform in this work has been recently advocated in the case of {\em bulk} correlators, in \cite{Hohler:2011wm},
in order for the corresponding spectral densities to satisfy appropriate sum rules. Here we are able to justify why this subtraction is physically
sensible, in the case of the Langevin process, but we expect that similar arguments can be used for the bulk spectral functions.

The need to define dressed correlators stems essentially from the need to cure a UV behavior which is too singular. On the other hand, we should
expect that this does not affect the {\it low-frequency} regime, which gives the transport coefficients and should be independent of the details of
how we treat short-distance physics. In other words, the subtracted correlators should have the same small frequency limit as the unsubtracted ones.

In this paper therefore we perform the following task

\begin{enumerate}

\item We define the dressed holographic Langevin evolution as described above, by performing the appropriate redefinitions in the path integral, and
we show that they have the correct fall off at high frequency for the dispersion relations to hold.

\item We analyze the low-frequency asymptotics of the zero temperature retarded force correlators in order to assess to what extent the dressing
affects IR asymptotics of the diffusion.

\item We compute numerically the dressed correlators $G_i(\omega,T)-G_i(\omega,0)$ both in $AdS$, and in exact scaling backgrounds as particular
examples, and verify that the dressed correlator has the desired properties. We check explcitly that the dressed correlators have compact support and
exhibit a few oscillations before they die-off beyond the natural scale set by $T_s$.

\end{enumerate}

In this paper we analyze the IR asymptotics of the retarded correlator of the force exerted on the fundamental string by the plasma as a function of
the IR asymptotics of the dilaton potential. While the IR asymptotics are irrelevant for the large frequency behavior of the spectral densities, they
are the feature that determines the diffusion constants. In this work we limit the analysis to models that give rise to zero-temperature geometries
that are {\it non-confining}. In this case, the identification of the trailing string solution at zero-temperature is straightforward, and the
dressing procedure can be defined consistently. The case of confining zero-temperature theories is of course more interesting , since it contains QCD,
but it is also more subtle, and will be analyzed separately, \cite{langevin-confining}.

The structure of this paper is as follows.

In Section \ref{review} we review the Langevin dynamics of a relativistic point particle and its holographic description. We give general expressions
for the large frequency behavior of the spectral densities, and we discuss some explicit examples. We extend the results found in \cite{langevin-1} to
non-asymptotically $AdS$ scaling backgrounds.

In Section \ref{vac} we compute the vacuum Langevin correlators, both in the high and low-frequency behavior, calculate the diffusion coefficients in
various types of IR geometries, and give a classification of the IR geometries accordingly.

In Section \ref{renormalized} we define the subtraction procedure from a fundamental perspective, using a path integral formulation of the Langevin
dynamics. We show that the dressed correlators defined in this way arise naturally if one imposes some basic physical requirements. In this Section we
also show that they have the correct fall-off behavior at large frequency to allow a consistent application of the dispersion relations.

In Section 5 we present numerical results for the dressed correlators in both $AdS$ and the non-conformal scaling geometries.

Some technical details are left to the Appendices.

\section{Review of the Langevin evolution in strongly coupled plasmas}
\label{review}

In this Section we review the description of generalized Langevin dynamics in holographic finite-temperature plasmas, and the corresponding result
\cite{langevin-1} for the high-frequency behavior of the spectral density. These results are extended in the last Subsection to non-asymptotically
$AdS$ scaling backgrounds.

\subsection{Generalized Langevin dynamics of a relativistic particle}

A probe heavy quark propagating in a deconfined plasma undergoes a generalised Langevin process, due to the interaction with the medium. If we assume
the quark to be heavy (compared to the temperature), we can instantaneously describe the fluctuations $\delta X^i(t)$ of the quark position with
respect to a straight trajectory, by an action of the form: \be\label{bound action} S[\delta X(t)] = S_0[\delta X(t)] + \int d\tau ~\delta X_\mu(\tau)
{\cal F}^\mu(\tau) \ee where $S_0$ is the free particle action and $ {\cal F}^\mu(\tau)$ is a field (operator in the operator formalism) which depends
only on the medium degrees of freedom and dynamics, and represents the effect of the microscopic interactions with the plasma. Then, the generalised
Langevin process is given by the equation:
\be
P^i(\delta X) = \int_{-\infty}^{+\infty} d\tau G_R^{ik}(\tau) \delta X_k(t-\tau) \,+\, \xi^i(t).
\label{lange}\ee
On the left hand side, $P^i(\delta X)$ represents the classical equation of motion for $\delta X^i(t)$ in the absence of dissipation, derived from
$S_0[\delta X]$. In the absence of external forces, $P^i(\delta X)$ typically takes the form of a two-time-derivative term:
\be\label{mass1}
 P^i(\delta X) = M_{eff}^{ij} \delta \ddot X_j,
\ee
where $M_{eff}^{ij}$ is an effective mass matrix. For a non-relativistic particle we simply have $M_{eff}^{ij} = M_Q \delta^{ij}$, whereas in the
case of a relativistic particle with velocity $\vec{v}$, it is given by (see e.g. \cite{langevin-1}):
\be\label{mass}
M_{eff}^{ij} = M_Q\left[\gamma^3(v) {v^i v^j \over v^2} + \gamma (v) \left(\delta^{ij} - {v^i v^j \over v^2}\right) \right], \quad \gamma(v) \equiv
(1-v^2)^{-1/2}.
\ee
$M_Q$ being the quark mass.

The effective force on the right hand side of (\ref{lange}) is composed of two terms:
\begin{itemize}
\item
an overall friction term, described by the convolution with a memory kernel given by the retarded correlator of the force operator ${\cal F}(t)$:
\be
G_R^{ij}(t) \equiv -i \theta(t) \left\< \left[{\cal F}^i(t) , {\cal F}^j(0)\right] \right\>,
\ee
\item a stochastic gaussian external force with moments given by the symmetrized correlator of the same operator ${\cal F}$:
\bea
&&\< \xi^i (t) \> =0, \nonumber \\
&& \< \xi^i(t) \xi^j(0) \> = G^{ij}_{sym}(t) \equiv -{i\over 2} \left\< \left\{{\cal F}^i(t) , {\cal F}^j(0)\right\} \right\>.
\eea
\end{itemize}
It is important to stress that the expectation values are calculated in the appropriate ensemble. In the non-relativistic limit, this is the plasma ensemble (heat bath). However at relativistic speeds the ensemble is different from that of the plasma ensemble, and is characterised by an emerging temperature $T_s$ different from that of the plasma, \cite{iancu,langevin-1}.

Going to Fourier space, one defines the {\em spectral density} by
\be\label{rtcorr}
\rho^{ij}(\omega) = \int_{-\infty}^{+\infty} dt \, e^{i\omega t} \left\< \left[{\cal F}^i(t) , {\cal F}^j(0)\right] \right\>.
\ee
One then has the dispersion relations:
\be \label{dispersion}
{\rm Im}\, G_R^{ij}(\omega) = -\pi \rho^{ij}(\omega) , \qquad {\rm Re}\, G_R^{ij}(\omega) = P\int_{-\infty}^{+\infty} d\omega' {\rho^{ij}(\omega') \over \omega - \omega'}
\ee
Note also here that these dispersion relations are valid only for appropriately subtracted densities, and we will expand on this below.

The retarded correlator does not depend of the specific density matrix of the medium, but only on the canonical commmutation relations (or OPE) of the operators ${\cal F}$ (see e.g. \cite{teaney} for a recent discussion).
On the other hand, by specifying the density matrix one obtains a relationship between the retarded and symmetric correlators. Thus, the spectral density plus a choice of ensemble determine completely the generalized Langevin equation. For example, in the case of a thermal ensemble in equilibrium at a temperature $T$, one has:
\be
G_{sym} = \pi \coth (\omega/2T)\, \rho(\omega).
\ee

The crucial consideration that motivates the present paper is that, given a spectral density $\rho(\omega)$, in order for the real-time correlators to
be well defined through equation (\ref{rtcorr}), and for the dispersion relations (\ref{dispersion}) to be valid, $\rho(\omega)$ must have a sufficiently fast fall-off at large $\omega$. If this is not the case,
 one has to perform a suitable subtraction in order to have a well defined Langevin process. As usual, a bad large frequency behavior is related to a too singular behavior of the real-time correlators in the coincident time limit.

That this is precisely the case in our situation can be seen heuristically by dimensional analysis: from equation (\ref{bound action}), the force operator ${\cal F}(t)$ has mass dimension 2, therefore one expects that for short time and large frequencies one has, in a CFT:
\be
 \rho(\omega) \sim \omega^3.
\ee
In coordinate space this implies a correlator with a delta-function singularity at the origin, ${\rm Im} \,G_R(t) \sim \delta'''(t)$.

The holographic calculation shows indeed that the unsubtracted spectral density grows in fact like $\omega^3$ in the $M_Q \to \infty$ limit, as we
will review shortly\footnote{For finite $M_Q$, it grows instead linearly in $\omega$, for $\omega \gg M_Q$. However, as we will argue below, this
regime is unphysical, since in this case the holographic picture of a heavy quark as the endpoint of the string with boundary is inconsistent.
Therefore, by ``large frequency'' we will always mean $\omega \gg T$, and for finite quark mass, one should always restrict the analysis to $T\ll
\omega \ll M_Q$.}. With this behavior, the Fourier integral (\ref{rtcorr}) and the dispersion relations (\ref{dispersion}) are ill defined.

The short-time divergence is determined by the UV behavior, which may be completely distinct from the dynamics of the medium in which the quark propagate. In the following sections we will define a subtracted spectral density, which physically means that the Langevin
dynamics will describe a ``dressed'' quark, in which the effect of the vacuum contribution has been renormalized away.

\subsection{Langevin correlators in holographic plasmas}

In gauge/gravity duality, a heavy external quark moving through the plasma at temperature $T$ can be described by a string whose endpoint at the
boundary follows the quark's trajectory \cite{her}-\cite{gub2}. The string extends into the bulk, whose geometry is a black hole background with
appropriate temperature $T$. The bulk fields $\delta X^i(r,t)$, describing the fluctuation of the trailing string around its classical solution,
correspond holographically to the force operator on the boundary. Thus, we can extract the force correlators of the previous section from the behavior
of the trailing string fluctuations, using the standard holographic rules.

Here, we will consider five-dimensional planar black holes, with metric\footnote{This is in the string frame. In the whole of this paper the metric will always be in the string frame. For a conformal theory this does not make a difference as the dilaton is constant. But for the non-conformal theories we will consider where the running scalar is the dilaton, it makes a big difference. }:
 \be\label{metric}
ds^2 = b^2(r)\left[{dr^2\over f(r)} - f(r)dt^2 + dx^i dx_i
\right].
\ee

The trailing string dual to the quark is governed by the Nambu-Goto action,
\begin{equation}\label{NGACTION}
S_{NG} = -\frac{1}{2\pi \ls^2}\int d^2\sigma \sqrt{-\det g_{\alpha\beta}} \;, \qquad g_{\alpha\beta}=
g_{\mu\nu} \partial_{\alpha}X^{\mu} \partial_{\beta}X^{\nu}, \qquad \left\{\begin{array}{c} \mu,\nu =0\ldots5\\
\a,\b= 0,1\end{array}\right.
\end{equation}
where $g_{\mu\nu}$ are the components of the bulk metric (\ref{metric}). We choose the gauge $\xi^0=t, \xi^1 =r$ for the world-sheet coordinates, and consider a classical string embedding of the form
\begin{equation}
\label{TRAILANSATZ}
 \vec{X}(t,r) = \left(v t + \xi(r) \right){\vec{v}\over v},
\end{equation}
i.e. such that the endpoint $r=0$ moves at constant velocity $\vec{v}$, and the rest of the string trails along in the bulk. The induced 2d metric on
the world-sheet has a horizon at $r = r_s$, defined by
\be
f(r_s) = v^2,
\ee
with associated Hawking temperature $T_s$:
\be
 T_{s} \equiv {1\over 4\pi}\sqrt{f(r_s)f'(r_s)\left[{4b'(r_s)\over b(r_s)}+{f'(r_s)\over f(r_s)}\right]}.
\ee

The fields dual to the force operator $\vec{{\cal F}}(t)$ are the fluctuations around the trailing string solution \ref{TRAILANSATZ}. It is necessary to distinguish fluctuations longitudinal and transverse to the background velocity $\vec{v}$. The fluctuating trailing string is described by the embedding
\be\label{FLUCANSATZ}
 \vec{X}(t,r) = \left( v t + \xi(r) + \delta X^\parl(t,r)\right){\vec{v}\over v} + \delta \vec{X}^\perp(t,r) , \quad \vec{v} \cdot \delta \vec{X}^\perp =0
\ee
where $\xi(r)$ is determined by the classical NG action. Expanding the Nambu-Goto action to second order we obtain the quadratic action governing the
fluctuations. In a convenient coordinate system, where the world-sheet metric is diagonal this takes the simple form:
 \be S_2=-{1\over 2}\int d\tau dr \left[ {\cal G}^{\a\b}_\parl
\partial_{\a}\dx\partial_{\b}\dx + \sum_{i=1}^2 {\cal G}^{\a\b}_\perp \partial_{\a}
\delta X_i^\perp\partial_{\b}\delta X_i^\perp \right]
\label{H16}\ee where the kinetic operators are defined by

\be \label{Gab} {\cal G}^{\a\b}_\perp \equiv {1 \over 2 \pi
\ell_s^2} H^{\a\b}, \qquad {\cal G}^{\a\b}_\parallel \equiv {1 \over 2
\pi \ell_s^2} {H^{\a\b} \over Z^2}, \ee
with
 \bea
 &&H^{\a\b} = \left( \begin{array}{cc}
-{b^4\over \sqrt{(f-v^2)(b^4f-C^2)}}~~ & ~~0\\
0~~ & \sqrt{(f-v^2)(b^4f-C^2)}\end{array}\right)\;, \label{Hab} \\
&& C = v b^2(r_s), \qquad Z= b^2\sqrt{f-v^2\over b^4f-C^2}. \label{H10}
\eea

For a harmonic ansatz of the form
$\delta X^i(r,\tau)=e^{i\omega \tau}\delta X^i(r,\omega)$,
the equations following from the action (\ref{H16}) are:

\be
\partial_r\left[ R \,\,\partial_r\left(\delta X^{\perp}\right)\right]+{\omega^2b^4\over R}\,\delta X^{\perp}=0,
\label{H19}\ee
\be
\partial_r\left[{1\over Z^2} R\,\,\partial_r\left(\delta X^{\parallel}\right)\right]+{\omega^2b^4\over Z^2 R}\delta X^{\parallel}=0,
\label{H20}\ee
where
\be\label{R}
R \equiv \sqrt{(f-v^2)(b^4f-C^2)}.
\ee

The holographic prescription for the retarded correlator,
 computed with the diagonal induced metric (\ref{Hab}), is given by:
\be\label{full GR}
G_R(\omega) = - \left[\Psi_R^*(r,\omega) {\cal G}^{rr} \pa_r \Psi_R(r,\omega)\right]_{\rm boundary} \;.
\ee
Here $\Psi_R(r,\omega)$ denotes collectively the fluctuations
$\delta X^\parallel$, $\delta X^\perp$, solutions of equations (\ref{H19}-\ref{H20}), and the factor
$ {\cal G}^{rr} $ is the appropriate one from equation (\ref{Gab}). The solutions $\Psi_R(r,\omega)$ must obey the
appropriate boundary conditions: unit normalization at the UV boundary, and in-falling at the world-sheet horizon, i.e
\be
\Psi_R(r,\omega) \simeq \Psi_h (\omega) \,\,(r_s-r)^{-{i \omega\over 4 \pi T_{s}}} \qquad r\sim r_s. \label{psihor}
\ee

The UV boundary is taken to be $r=0$ in case we consider the quark mass infinite, or $r=r_Q>0$ for a finite quark mass $M_Q$. The value $r_Q$ is determined by calculating the mass as the Nambu-Goto action of a straight string extending from $r_Q$ down to the deep IR region. For a heavy quark, $r_Q$ is close to the boundary and $M_Q \sim 1/r_Q$.

The spectral density is related by equation (\ref{dispersion}) to the imaginary part of equation (\ref{full GR}). Since the latter is a conserved flux, one can evaluate it at the horizon $r_s$ rather than at the boundary. The result is:
\be\label{IMGR}
 \rho^\perp(\omega) = {b^2(r_s) \over 2\pi^2 \ell_s^2} \,\omega\,|\Psi_h^\perp(\omega)|^2 \qquad \rho^\parl(\omega)= {b^2(r_s) \over 2\pi^2 \ell_s^2 Z^2(r_s)} \, \omega\,|\Psi_h^\parl(\omega)|^2 ,
\ee
where $\Psi_h(\omega)$ are the coefficients of the in-falling wave-functions, see equation (\ref{psihor}).

Equation (\ref{IMGR}) shows that once the background metric is known, the spectral density is determined by the coefficient that governs the horizon
asymptotic of the in-falling fluctuation wave function.

\subsection{The high-frequency limit}\lab{high freq}

In most of the following we assume there is an asymptotically $AdS$ region $r \to 0$
where
\be\label{UVmetricdil} \log b(r) \sim -\log {r\over \ell} +
{\rm subleading} , \qquad f(r) \sim 1 + O(r^4), \qquad r\to 0, \ee and a
horizon at $r=r_h$ where $f(r_h) = 0$, and $f'(r_h)$ and $b(r_h)$
remain finite. The black hole temperature is given by:
\be 4\pi T =
-f'(r_h). \ee
Later we will generalize some of the results to other backgrounds that do not satisfy (\ref{UVmetricdil}), that will be discussed separately in Section \ref{scaling}.

We make no particular assumptions on the subleading terms in equation
(\ref{UVmetricdil}). In case these subleading terms actually vanish sufficiently fast as $r\to 0$, then the metric is asymptotically $AdS$ in the usual sense.

In \cite{langevin-1} we determined the high-frequency behavior of the spectral density from equation (\ref{IMGR}) by approximately solving eqs. (\ref{H19}) and (\ref{H20}) for large $\omega$. This can be done using an adaptation of the WKB method, as we review in Appendix \ref{wkbapp}. Here, by high-frequency limit, we mean the limit $\g \omega \gg T,\Lambda$ (where $\Lambda$ is the scale that drives the breaking of conformal symmetry, and can be linked to a deformation by a relevant or a marginally relevant operator in the UV). More precisely, the WKB calculation in \cite{langevin-1} shows that the high-frequency approximation
holds for $\gamma \omega r_s \gg 1$. Since $r_s \simeq 1/(\sqrt{\g}T)$ this implies $ \sqrt{\g}\omega \gg T$.

 On the other hand, even for finite quark mass $M_Q$, we will always restrict $\g \omega \ll M_Q$. The reason is that, in the opposite limit, the description of the heavy quark we are using becomes inconsistent. This is due to the fact that for large frequency of the fluctuations the accelerations become large, while we are assuming here that the configuration is a small perturbation around a steady $v=const$ motion. For large $\g\o \gg M_Q$ the backreaction of the fluctuations cannot be neglected. Another, related problem is that for $\g \o \gg M_Q$, the boundary on-shell action for the fluctuations does not reduce to the kinetic action for a quark of mass $M_Q$ (as it is the case in the opposite limit, $M_Q \gg \g \o$) but rather it gives a non-derivative potential term. Thus, the holographic pictures of the heavy quark seems to be invalid for modes of arbitrarily high frequency.

 To summarize, the interesting high-frequency regime, for a given temperature, quark mass, and quark velocity, is\footnote{For this interval to be nonempty one needs $r_Q < r_s$, otherwise the trailing string picture is inconsistent even at the level of the background solution.}:
\be \label{regime}
 {1 \over r_s} \ll \g\o \ll {1\over r_Q} \qquad \Leftrightarrow \qquad {T\over \sqrt{\g}} \ll \o \ll {M_Q \over \g }.
\ee
In this regime, the spectral densities are approximated by \cite{langevin-1}:

\be\lab{Moo}
\rho_{\perp}(\o) \simeq \gamma^{-2}\rho_\parl (\o) \simeq {\gamma^3 \over 2\pi^2\ell_s^2} \,\o^3\,\, \frac{ r_{tp}^2(\omega) R_{tp}(\omega)}{ 1 + (\gamma \o r_Q)^2 + O\left((\g\o r_Q)^4\right)}.
\ee
The derivation of this expression is presented in in Appendix \ref{wkbapp}. Here, $r_{tp}(\omega)$ is the classical turning point of the differential equation for the fluctuations, $R_{tp}(\omega) \equiv R(r_{tp}(\omega))$, with the function $R(r)$ defined in equation (\ref{R}). This general expression simplifies when we take into account
 the explicit asymptotic form of $R(r)$ and when we take $M_Q \to \infty$, as we will see below. Here, we make a few important remarks.
\begin{itemize}
\item The expression (\ref{Moo}) gives the leading high-frequency behavior of the spectral densities in terms of the background metric. More specifically, the result is completely determined by the UV geometry close to the boundary.
The only assumptions needed to derive them are that, close to the boundary, the metric obeys relaxed $AdS$ asymptotics:
\be\label{relaxed}
b(r) \to { \ell\over r}h(r), ~~\;\; {\rm with}\;\;~~ {r h' \over h} \to 0 ~~\;\; {\rm as}\;\;~~ r\to 0.
\ee
Under this assumption, the classical turning point of the differential equation is approximately,
\be\label{rtp}
r_{tp} \simeq \frac{\sqrt{2}}{\gamma \o} \ll r_s
\ee
for large $\o$. Also, the boundary endpoint $r_Q$ scales approximately as $1/M_Q$ for large\footnote{Large compared to the temperature scale, and eventually the scale that breaks conformal invariance in the model.} $M_Q$. Thus for large $M_Q$ and large $\o$, the metric functions appearing in equation (\ref{Moo}) are evaluated at a point close to the $AdS$ boundary $r=0$.
\item For $r_{tp}$ and $r_Q$ close to the $AdS$ boundary, one can approximate
\be\label{R2}
R(r) \simeq {b^2(r) \over \gamma} = {\ell^2 \over r^2} {h(r)\over \gamma}.
\ee
which easily follows from equation (\ref{R}).
\item
In the limit $M_Q\to \infty$, i.e. when the quark is considered non-dynamical, the trailing string description is valid for arbitrarily large frequency. We can then drop the denominator, and replace (from equations (\ref{rtp}-\ref{R2})).
$
r_{tp}^2 R (r_{tp}) \sim r_{tp }^2b^2(r_{tp})/\gamma \sim \ell^2 h^2(\sqrt{2}/\gamma\o),
$
\end{itemize}
Then, equation (\ref{Moo}) simlpifies to:
\be\label{Moo2}
\rho_{\perp}(\o) \simeq \gamma^{-2}\rho_\parl (\o) \simeq {\ell^2 \over 2\pi^2\ell_s^2}\gamma^2 \o^3\,\,h^2(\sqrt{2}/\gamma\o).
\ee
This equation shows that, in asymptotically $AdS$ backgrounds, and for $M_Q \gg \gamma \o$, the spectral density behaves approximately as $\o^3$ at large frequency, as anticipated.
The leading cubic behavior is deformed by the extra $\o$-dependence induced by $h(1/\o$),
which due to the condition (\ref{relaxed}), has a milder dependence than any power-law for large $\o$.
The function $h(\o)$ encodes the deviation from conformal invariance, and it may also contain a temperature dependence.

In the following Subsections we will describe two explicit examples: the conformal case (i.e. a pure $AdS$-Schwarzschild background), and the
logarithmically-deformed $AdS$ asymptotics one finds in Einstein-Scalar theories like the Improved Holographic QCD setup discussed in
\cite{langevin-1}. This last example captures also the cases that involve relevant perturbations in the UV. Finally, we give the generalization of
equation (\ref{Moo2}) for backgrounds that exhibit scaling but are not asymptotically $AdS$.

\subsubsection{Conformal diffusion}\lab{conformal diffusion}
When the dual field theory is conformally invariant, the 5D background is an $AdS$ black hole, whose metric is given by equation (\ref{metric}) with:
\be\lab{AdS metric}
b(r) = {\ell\over r}, \qquad f(r) = 1 - (\pi T r)^4,
\ee
where $\ell$ is the $AdS$ length and $T$ the bulk Hawking temperatures. For a boundary quark moving at speed $v$, the world-sheet black hole temperature $T_s$ and the functions $Z$ defined in equation (\ref{H10}) are respectively:
\be
T_s = {T \over \sqrt{\gamma}}, \qquad Z = {1\over \gamma}.
\ee
Since $Z$ is independent of $r$, the transverse and longitudinal fluctuations obey the same equation (see equations (\ref{H19}-\ref{H20}), and the only difference between $\rho^\perp$ and $\rho^\parl$ comes from the boundary normalization, i.e. $\rho^\parl = \gamma^2 \rho^\perp$.

The quark mass $M_Q$ determines the boundary endpoint of the trailing string through the relation:
\bea\lab{mass-cutoff}
M_Q = {\ell^2 \over 2\pi \ell_s^2 }{1 \over r_Q}.
\eea

The fluctuation equation for both transverse and longitudinal modes only depends on the dimensionless variables $x\equiv r/r_s$ and $\tilde \omega
\equiv {\omega\over 4\pi T_s}={\omega \sqrt{\gamma}\over 4\pi T}$:
\bea\label{fluctu conf}
\partial_x \left[ \frac{1-x^4}{x^2} \partial_x \Psi(x,\tilde\omega) \right] + \frac{(4 \tilde \omega)^2}{x^2 (1-x^4)} \Psi(x,\tilde\omega) = 0.
\eea

Let us now specify the general results (\ref{Moo}) to the present situation. In the UV limit $r\to 0$, we have $R \sim \gamma^{-1}\ell^2/r^2$, and
$h(r) =1$. From (\ref{Moo}) we obtain the following large-$\o$ asymptotics: 
\be\label{conformal}
\rho^\perp (\o) = \gamma^{-2} \rho^\parl (\o) \simeq {\ell^2 \over 2\pi^2 \ell_s^2}\, \gamma^2 \,\o^3 
\ee
In the equations above, one can substitute the ${\cal N} = 4$ relation between $\ell_s$ and $\ell$, $(\ell/\ell_s)^2 = \sqrt{\lambda_{{\cal N}=4}}$.

The result (\ref{conformal}), to leading order in the frequency, will hold in any geometry which is asymptotically $AdS$ in the strict sense, i.e. any
geometry in which the scale factor in the string frame behaves strictly as $\ell/r$ for $r\to 0$, i.e. as in \refeq{relaxed} with $h(r)= 1 +\cO(r) $ for
$r\to0$
(In Einstein-dilaton models, this means that the dilaton has to be constant in the UV if the Eintein frame scale factor is asymptotically $AdS$). In these geometries the
deviation from conformality will give rise to subleading corrections in inverse powers of $\omega$
with respect to the result (\ref{conformal}).
In the next Subsection we will consider an example
where this is not the case, because of the presence of a non trivial dilaton (hence $h$ does not go to $1$ in the UV),
and breaking of conformal invariance in the UV deforms the leading $\o^3$ behavior.

\subsubsection{Running dilaton backgrounds} \lab{IHQCD}

A setup which exhibits breaking of conformal invariance in the UV in a manner that mimics QCD can be constructed from the 5D Einstein-dilaton theory with action in the Einstein frame \cite{ihqcd1,ihqcd2}:
 \be\label{action2} S =
-M_p^3 N_c^2 \int \sqrt{-g^E} \left[R^E - {4\over3} {(\nabla
\lambda)^2\over \lambda^2} + V(\lambda)\right]. \ee In the
holographic interpretation of these models, the scalar $\lambda$
is dual to the running coupling $\l_t$ of the four-dimensional
gauge theory.

This type of actions provides a general class of holographic dynamics, describing a CFT perturbed by a relevant scalar operator.
In the conventional case of relevant perturbations, setting $\phi=\log\l$ and adjusting the UV fixed point to be at $\phi=0$, the near UV potential behaves as $V={12\over \ell^2}\left[1+{\cal O}(\phi^2)\right]$.

For a theory with a UV fixed point at $\l=0$, the potential should have a regular
expansion as $\lambda \to 0$, with
\be V(\lambda) \sim {12\over \ell^2}(1
+ v_0 \lambda + \ldots)\,\,.
\label{pot}\ee
With these requirements, the solutions in the Einstein frame are an asymptotically $AdS$ metric, with $AdS$ length $\ell$, and a non-trivial profile $\lambda(r)$ which vanishes logarithmically at the $AdS$ boundary $r=0$:
\be
\label{UVlimit} b_E(r) \sim {\ell\over r} \left[1 + O\left(1\over \log r\right) \right],
\quad \lambda(r) \sim -{9\over 8 v_0 \log r} + O\left({\log \log r \over \log^2 r}\right), \qquad r \to 0.
\ee
The scale factor in the above equation is written in the Einstein frame. On the other hand, the geometry (\ref{metric}) felt by the trailing string is the one in the string frame, which in a non-trivial dilaton background is given by
\be\label{stringmetric}
b(r) = \lambda^{2/3} (r) b_E(r)
\ee
Thus the string frame metric is not asymptotically $AdS$, but it nevertheless obeys the relaxed requirement (\ref{relaxed}), with
\be\label{h}
h(r) = \lambda^{2/3}(r) \left[1 + O(\lambda(r))\right] \sim \left(-{9\over 8v_0\log r}\right)^{2/3}[1 + \ldots]
 \ee
 For an appropriate choice of the potential $V(\l)$, the models with action (\ref{action2}) and asymptotics as in (\ref{pot})
provide a good holographic dual to large-$N_c$ 4-dimensional pure
Yang-Mills theory, at zero and finite temperature
\cite{ihqcd1}-\cite{gkmn3}.

For a short review of the main features of these models, the
reader is referred to \cite{ihqcdrev}.

We now consider the Langevin spectral densities in this class of models.
Using equation (\ref{h}) we obtain from (\ref{Moo2}) the large-$\o$ behavior:
\be\label{non-conformal}
\rho^\perp(\o) \simeq \gamma^{-2} \rho^\parl (\o)
 \simeq \left(9\over 8 v_0\right)^{4/3}{\ell^2 \over 2\pi^2 \ell_s^2} \,\,{\gamma^2 \omega^3 \over (\log \omega)^{4/3}}\left[1 + O\left(1\over \log \o\right)\right].
\ee
As for the conformal case, we remark that the above results are temperature independent.

The leading result $\sim \o^3/(\log \o)^{4/3}$ in (\ref{non-conformal}) is corrected by a power series in $\lambda_{tp}~(\log \o)^{-1}$, which stems from the fact that in deriving equation (\ref{non-conformal}) we have ignored terms in the fluctuation equation that are proportional to $r \l'/\l$ \cite{langevin-1}, and also from the series expansion in $h(r)$. From the background Einstein equation these terms can be expressed in a power series in $\lambda$ using the superpotential function $W(\l)$ (see \cite{gkmn2}):
\be
r \l'/\l = \l W(\l) = \sum_{n=0}^{\infty} W_n \l^{n+1}
\ee
where the coefficients in the power series are determined by the series expansion coefficients of $V(\l)$. Thus, the leading result
(\ref{non-conformal}) will have the form of a series in $\lambda_{tp}$ (which is a small parameter for large $\omega$), i.e.
\be
\rho^\perp(\o) \propto {\omega^3 \over \log^{4/3}\o} \left[1 + \sum_{n=1}^{\infty} {c_n \over\log^n \o} \right]
\ee
The coefficients $c_n$ can in principle be determined order by order from the series expansion of the superpotential. {\em These coefficients are
temperature-independent}, as shown in \cite{gkmn2}. Thus, although the leading $\o^3$ term is corrected by an infinite power series in inverse
logarithms, the temperature dependence cannot enter at any order in this series, and it is thus confined to smaller power-law corrections in $\o$.

\subsubsection{Scaling solutions}\label{scaling}
The result (\ref{conformal}) can be generalized to solutions that are not asymptotically $AdS$ black holes, but are such that the string frame scale factor and blackness function behave as simple power laws:
\be\label{scbk}
b(r) = \left({\ell \over r}\right)^{\bar{a}} , \qquad f(r) = 1 - \left({r\over r_h}\right)^c
\ee
These backgrounds are solutions of the Einstein-Dilaton system in $p+1$ dimensions, if the scalar potential is a simple exponential (see Appendix \ref{scalingapp} for details):
 \be\label{action3} S =
-M_p^3 \int \sqrt{-g} \left[R - {1\over2} (\de\phi)^2 + V(\phi)\right], \qquad V(\phi) = 2 \Lambda e^{-\delta \phi}. \ee

The interest in these backgrounds is that they arise
in the IR behavior of all Einstein-dilaton theories with a potential that asymptotes to a single exponential at large $\phi$, \cite{cgkkm}.
They also describe non-conformal branes in string theory, and their holographic renormalization is well understood \cite{skendi}. They can be obtained by dimensional reduction of $AdS$ black holes in a higher-dimensional pure gravity theory, and this explains their scaling properties in terms of a ``hidden'' higher dimensional conformal symmetry \cite{GK}.

The second order equation satisfied by the fluctuation is
\be
x^{2\bar a}\partial_x\left[x^{-2\bar a}\sqrt{(1-x^c)(1-(1-v^2)x^c-v^2x^{4\bar a})}\pa_x\left(\delta X^{\perp}\right)\right]
\label{23a}\ee
$$
+{\Omega^2 \over \sqrt{(1-x^c)(1-(1-v^2)x^c-v^2x^{4\bar a})}}\delta X^{\perp}=0
$$
with
\be
\Omega={w_s~\omega\over \sqrt{1-v^2}}={c(1-v^2)^{2-c\over 2c}\over 4\pi}~{\omega\over T}
\label{24a}\ee
It is derived in appendix \ref{scalingapp}. $x$ is a scaled radial coordinate and the various constants appearing in the equation are defined in the appendix \ref{scalingapp}.

The high-frequency analysis of the world-sheet fluctuations in these backgrounds leads to the following spectral density (as we show in detail in Appendix \ref{wkbapp}):
\be\label{rhoscaling}
\rho^\perp \simeq \gamma^{-2}\rho^\parl \simeq {\ell^{2\bar{a}} \over \pi \ell_s^2} {\g^{2\bar{a}}\over 2^{2\bar{a}+1} \Gamma^2(\bar{a}+1/2)} \o^{2\bar{a}+1} \left[ 1 + O\left((\g\o r_Q)^2\right)\right].
\ee
The case $\bar{a} = 1$ (which arises when $\delta=0$ in equation (\ref{action3}) correspond to the $AdS$ scaling $\sim \o^3$, and (\ref{rhoscaling}) reduces to the conformal result (\ref{conformal}).
\vskip 1cm
The results in this section all lead to spectral densities that have a divergent large-$\o$ behavior, in particular
$\rho(\omega) \propto \sim \o^3$ for conformal or asymptotically conformal theories. As discussed in Section 2.1, this cannot be the correct
behavior for the {\em physical} spectral densities that are used to compute real-time Langevin correlators, since they cannot satisfy the
appropriate dispersion relations (\ref{dispersion}). For these relations to hold, $\rho(\o)$ should have a well defined Fourier transform, and
it should decay at least as $1/\o$ at large frequencies.

To put it differently, the spectral densities we found above would imply that dissipative
and retardation effects would be stronger for very high energy modes. On the other hand, on physical grounds we should expect that
the high-frequency modes would behave like in the vacuum, i.e. obey a classical equation with no retardation and no noise term.

As we will show in the next Section, the resolution of these problems becomes clear once we realize that, even in the zero-temperature vacuum,
the spectral densities computed holographically are non-zero, and display the same large-frequency behavior.
This suggest that spectral densities satisfying physical requirement at any temperature $T$ can be obtained by subtracting
the vacuum contribution. In Section \ref{renormalized} we will show that this is indeed the correct prescription. Before doing that, we
need to compute the vacuum spectral densities.

\section{The vacuum force correlators\label{vac}}

The zero temperature correlators for an infinitely massive quark moving in the conformal plasma have been analytically derived in \cite{gubser}. In this
Section we generalize those results to a generic $T=0$ background with metric given by \refeq{metric}, but with $f(r)\equiv1$. We will also compute
the spectral density in the case of a quark with finite (but large) mass, $M_Q \gg \g \o$ (as explained in Section \ref{high freq}), both in the
conformal background and in non conformal backgrounds, such as the scaling backgrounds defined by \refeq{scbk}.

In this paper, we restrict to backgrounds such that the string frame scale factor $b(r)$ is a {\it monotonically decreasing} function of $r$ (from the
UV to the IR). This means that the gauge theory side at zero-temperature is not confining. In this situation, there is no conceptual problem in
describing the propagation of a single quark through the zero-temperature medium. On the other hand, for backgrounds dual to confining gauge
theories, this is not evident, and one runs into subtleties. As we will see in Section \ref{confiningsec}, these subtleties will result in some
immediate problems if we simply try to generalize the results obtained in this Section to the confining case. The proper treatment of the dressed
Langevin dynamics in the confining case will be given in a separate work \cite{langevin-confining}.

Taking the trailing string anzatz \refeq{TRAILANSATZ}, the configuration that minimizes the Nambu-Goto action \refeq{NGACTION} at zero temperature is
a straight string stretching from the boundary to the interior of the bulk, characterized by $\xi'(r)\equiv0$.

Considering the fluctuating trailing string described by the ansatz \refeq{FLUCANSATZ}, we can compute the action for the world-sheet fluctuations.
Using the trailing string embedding, $\xi'(r)\equiv0$, the Nambu-Goto action at second order in the fluctuations reads
\bea
S_{2} = - {1\over 2} \int dt\,dr \left[\cG^{\a\b}_\parl \dt_\a \d X^1 \dt_\b \d X^1 + \cG^{\a\b}_{\perp} \sum_{a=1}^2 \dt_\a
\d X^a \dt_\b \d X^a \right],
\eea
where $\cG^{\a\b}_\parl$ and $\cG^{\a\b}_\perp$ defined as in \refeq{Gab} with $Z(r)\equiv{1 / \g}$ and
\bea
H^{\a \b} = \g b(r)^2 \,\diag\left( -1, {1 \over \g^2} \right).
\eea

Hence the fluctuation equation at zero temperature becomes
\bea\lab{0T fluct eq}
\Psi''+ 2 {b' \over b} \Psi' + \g^2 \o^2 \Psi = 0,
\eea
both for transverse and longitudinal modes, $\Psi = e^{i\o t} \d X^i,$ $i=1,2,3.$

In the UV region $r\sim 0$ this fluctuation equation has the same form as the one for finite temperature.
On the other hand, in the IR there is no horizon, and the equation depends on the behavior of the scale factor at large $r$. We will restrict the attention to models for which the conformal coordinates extends to infinity, and such that the (string frame) scale factor has a power-like behavior in as $r\to +\infty$:
\bea\lab{power}
b(r)\sim {r}^{-a}\,, \quad r\to +\infty, \quad a>0
\eea
In particular, the conformal case corresponds to $a=1$, whereas in the Einstein-dilaton models of Section (\ref{IHQCD}) the exponent $a$ is determined by the behavior of the dilaton potential for large $\lambda$.
 A power-law fall-off of the scale factor as in (\ref{power})
corresponds to a dilaton potential that at large $\lambda$ behaves as
$V(\l) \sim \l^Q$ with $Q< 4/3$. With this parametrization the power $a$ is determined by the relations $a=(1-9Q^2/16)^{-1}$ if $\log \lambda$ is {\it not} the string dilaton, or $a = (1+3Q/4)^{-1}$ if it is \cite{ihqcd2}.

Using (\ref{power}) in the wave equation, we immediately obtain the following asymptotic expression for the in-falling wave function
\bea\lab{0T wave ir}
\Psi_R(r,\o)\simeq C_\o \, r^{a} e^{i \o r}\,, \quad r\to\infty,
\eea

As in Section 2, evaluating the conserved flux ${\rm Im} \Psi^* {\cal G}^{rr}\de_r \Psi$ at the horizon we obtain the spectral densities:
\be\label{0Tspectral}
\rho_o^\perp = \gamma^{-2} \rho^\parl_o = {\left[r^{a} b(r)\right]_{r\to \infty} \over 2\pi^2\ell_s^2} \o |C_\o|^2
\ee

The numerator is finite thanks to the asymptotic behavior (\ref{power}). Notice that, although the black hole scale factor $b(r)$ reduces to the zero-temperature scale factor as $r_h \to \infty$, equation (\ref{0Tspectral})
 is {\emph not} the limit of equation (\ref{IMGR}) for $r_s\to \infty$, which in fact vanishes as $\sim r_h^-a$.

We can use the WKB computation outlined in Appendix \ref{wkbapp} in order to extract the high-frequency behavior of the zero
temperature spectral densities. The only difference in the WKB computation at zero temperature is in the form of the in-falling wave-function, which in this case is given by (\ref{0T wave ir}). All other steps in the determinations of the coefficient $\Psi_o(\o)$ are unchanged, and we arrive at exactly the same expression as \refeq{non-conformal}. It now contains zero-temperature background quantities. With $f(r)=1$, $R(r) = b^2(r)/\gamma$, the spectral densities at zero temperature read, in the large $\o$ regime:
\be\lab{Moo0}
\rho^{\perp}_o(\o) \simeq \gamma^{-2}\rho^\parl_o (\o) \simeq {\gamma^2 \over 2\pi^2\ell_s^2} \,\o^3\,\, \frac{ r^2_{tp}(\omega) b^2_{tp}(\omega)}{ 1 + (\gamma \o r_Q)^2 + \cO\left((\g\o r_Q)^4\right)}.
\ee

Notice that the behavior is asymptotically $\o^3$ independently of the power-law
(\ref{power}) taken on by the scale factor in the IR. In other words, the
high-frequency limit, as expected, depends only on geometry in the UV region.
Moreover, the leading coefficient is exactly the same as at any non-zero $T$,
cfr. equation (\ref{Moo}).

Next, we evaluate the explicit large $\o$ behavior in the examples of interest.

\subsection{The conformal case}\label{0Tconf}

In the conformal case, one can solve equation \refeq{0T fluct eq} exactly, and evaluate equation (\ref{full GR}) directly in the UV, bypassing the expression (\ref{Moo0}). The exact solution for $\Psi(r)$ reads:
\bea\lab{conf wave}
\Psi(r) = A_1\le[ \cos (\g \o r) + \g \o r \sin(\g \o r) \ri] + \ A_2 \le[ \sin (\g \o r) - \g \o r \cos(\g \o r) \ri], \lab{conf 0T wave uv}
\eea
and the spectral densities are given in terms of $A_1$, $A_2$ by:
\bea\lab{0T rho}
\rho^{\perp}= \g^{-2} \rho^{\parl} = { \g^2 \o^3 \over 2 \pi^2 \ell_s^2}\, \im \le[ { A_1^* A_2} \ri]
\bigg|_{\rm boundary}.
\eea
We can immediately deduce the form of the exact spectral densities by setting the relation between $A_1$ and $A_2$ from the in-falling condition at
infinity \refeq{0T wave ir}:
\bea
A_1 = - i A_2.
\eea
The normalization at the cutoff boundary $r=r_Q$ implies
\bea
| A_1|^2 = |A_2|^2 = {1 \over 1 + \le(\g \o r_Q \ri)^2}.
\eea
From \refeq{0T rho} we get the exact result:
\bea\lab{conformal 0T}
\rho^{\perp}_o= \g^{-2} \rho^{\parl}_o = {\ell^2 \over 2 \pi^2 \ell_s^2} { \g^2 \o^3 \over 1 + \le( \g \o r_Q \ri)^2 }.
\eea
 In the limit of infinite mass $M_Q\to\infty$, or $r_Q \to 0$, this reproduces the formula for the spectral densities in \cite{gubser}, once we
use $\ell^2/\ell_s^2=\sqrt \l_{\cN=4}$.

\subsection{Running dilaton backgrounds}

Specialising equation (\ref{Moo0}) to the running dilaton models of Section \ref{IHQCD},
we obtain for the zero-temperature spectral densities in the large $\o$, infinite mass limit the same expression as in (\ref{non-conformal})
\be\label{non-conformal-2}
\rho^\perp(\o)_o = \gamma^{-2} \rho^\parl_o (\o)
 \simeq \left(9\over 8 v_0\right)^{4/3}{\ell^2 \over 2\pi^2 \ell_s^2} \,\,{\gamma^2 \omega^3 \over (\log \omega)^{4/3}}\left[1 + O\left(1\over \log \o\right)\right].
\ee

\subsection{Scaling backrgounds}\label{0Tscal}
For zero-temperature scaling backgrounds, described by equation (\ref{scbk}) with $f\equiv 1$, we can once again solve the vacuum fluctuation equations (\ref{0T fluct eq}) exactly, in terms of Bessel functions:
\be
\Psi(r) = \sqrt{\pi \over 2} e^{-i \frac\pi2 \bar a} C_\o \sqrt{\g \o r}\, r^{\bar a} \le[ J_{-\bar a-\frac12}(\g \o r) + i Y_{-\bar a-\frac12}(\g \o \r) \ri],
\ee
where $J_{\n}$ and $Y_{\n}$ are the Bessel function of the first and second kind respectively and the linear combination is chosen so that the
wave function in the IR is an in-falling wave.
Imposing unit normalization at the boundary $r=r_Q$, with $\g \o r_Q \ll 1$, we obtain the expression for the spectral density:
\be\lab{rho scaling}
\rho_o^\perp = \gamma^{-2} \rho_o^\parl = {\ell^{2\bar a} \over 2 \pi^2 \ell_s^2} {1 \over |c_{\bar a}|^2} {\g^{2\bar a} \o^{2\bar a+1} \over 1 + {(\g \o r_Q)^2 \over 2\bar a-1}},
\ee
for $\bar a>1/2$. Here $c_{\bar a}$ is the complex coefficient appearing in the leading order in the expansion of the Bessel function of order $\n = -\bar a-1/2$:
\be
|c_{\bar a}|^2 = {2^{2\bar a} \over \pi} \G^2\le( \bar a+\frac12 \ri)
\ee
The approximation in \refeq{rho scaling} is valid as long as $\g \o r_Q \ll 1$. It can be made more accurate by adding the next term,
$\cO\le(\o^{2\bar a+1}\ri)$, coming from the Bessel expansion. This term becomes more and more important as we send $\bar a$ to $1/2$. For half-integer values of
$a$ a logarithm appears in this term, making it of order $\le( \g \o r_Q\ri)^{2\bar a+1} \log(\g \o r_Q) $.

The exact expression for the spectral density for all values of $\bar a >0$ reads
\be\lab{rho scaling exact}
\rho_o^\perp = \gamma^{-2} \rho_o^\parl = {\ell^{2\bar a} \over \pi^3 \ell_s^2} {1 \over \g r_Q^{2\bar a+1}|J_{-\bar a-\frac12}(\g\o r_Q)+i
Y_{-\bar a-\frac12}(\g\o r_Q)|^2}.
\ee

For $\bar a \leq 0$ some pathologies can arise, as we will describe in Section \ref{confiningsec}. This class of models will be further analyzed in a
forthcoming paper \cite{langevin-confining}.

For $0<\bar a\leq 1/2$, the finite mass correction in the denominator of \refeq{rho scaling} is slower than $(\g \o r_Q)^2$. More precisely, it would give a
lower power behavior (or power times log, for $\bar a=1/2$), namely $(\g \o r_Q)^{2\bar a+1}$.

The results above, equation \refeq{rho scaling} or \refeq{rho scaling exact}, reduce to the conformal expressions of the previous Section if we substitute $\bar a=1$.

\vskip 1cm

As could have been expected, the leading UV asymptotics of the spectral functions is the same at finite and zero-temperature, and both display a growth as $\o^3$ (or $\o^{2\bar{a}+1}$) for scaling solutions).
 This suggests that we can obtain correlators that are well behaved in the UV, simply by subtracting the zero-temperature spectral densities from the finite temperature one. This method has been already applied in the holographic context for the bulk spectral densities \cite{Hohler:2011wm}. In the next section we will provide an {\it a priori} justification that this is in fact the correct prescription. There are
two points however that we need to be careful about, if we want to perform such a subtraction:
\begin{enumerate}
\item The $T=0$ subtraction must kill all terms that grow like $\o^3$ and $\o$, and leave at most an $\o^{-1}$ behavior at high frequency. In order for this to work, all divergent terms must have temperature-independent coefficients.
\item Subtracting the $T=0$ correlator should not modify the {\it small} frequency behavior at finite $T$, which controls the diffusion constants: it would be suspicious if the need to regulate a short-distance behavior would
also change the long-distance physics.
\end{enumerate}
The first point will be analyzed in detail in Section 4, where we explicitly construct the subtracted correlators. The second point is the subject of the next Subsection.

\subsection{The diffusion constant at zero temperature}\lab{diff const}

In this Subsection we will explicitly show that the zero temperature diffusion constant defined by
\bea\lab{kappa}
\eta = - \lim_{\o \to 0} {\im~G_R(\o) \over \o}
\eea
vanishes for all backgrounds with asymptotics given by \refeq{power}. As a particular case, we recover the vanishing of the zero temperature diffusion
constant in the conformal background, i.e. $a=1$. At finite
temperature, the computation leading to the Langevin diffusion constant was performed in detail in the Appendix B of \cite{langevin-1}, making use of
the formula \refeq{IMGR}. Since there is no horizon at $T=0$, the method must be modified, evaluating the wave function at $r\to\infty$,
rather than $r=r_s$.

First of all, we derive the expression for the diffusion constant in terms of the coefficient $C_{\o}$ appearing in the IR behavior of the
wave function, \refeq{0T wave ir}. Substituting the solution \refeq{0T wave ir} into the definition of the retarded correlator \refeq{full GR}, and
finally using \refeq{kappa}, we immediately obtain
\bea\lab{kappa ir}
\eta = \lim_{\o\to0} |C_\o|^2.
\eea

The next step is to compute $C_\o$. This can be done in analogy with the computation of the coefficient $\Psi_h$ defined in \refeq{psihor}, and
appearing in \refeq{IMGR}, in the finite temperature case (see \cite{langevin-1}). Determining $C_\o$ requires to match the IR and UV solutions of (\ref{0T fluct eq}).
in the small frequency limit.

In the IR, $b(r)$ is given by (\ref{power}) and the solution is
written in terms of Bessel functions,
\bea\lab{0T ir asympt}
\Psi(r) &\simeq& \sqrt{\pi \over 2} e^{-i \frac\pi2 a} C_\o \sqrt{\g \o r}\, r^{a} \le[ J_{-a-\frac12} (\g \o r) + i Y_{-a-\frac12} (\g \o r) \ri] \non\\
&\simeq& \le\{ \begin{array}{l@{\quad}l} 
C_\o \le( c_a\, \o^{-a} + d_a\, \o^{1+a} r^{1+2a} \ri)\,, & \o\to0
\\
~C_\o\le( 1 - \frac{i}{2} \frac{a+1}{\o r} \ri) r^{a} e^{i\o r} \,, & r\to\infty
\end{array} \ri.
\eea

 The (UV normalized) zero frequency solution of \refeq{0T fluct eq} on the other hand is given by

\bea\lab{0T 0omega}
\Psi_{\o=0}(r) = C_s + C_v \int dr \, b^{-2}
\simeq \le\{ \begin{array}{l@{\quad}l} 
C_s + {C_v \over 1+2a} \left(r^{1+2a}+ K\right)\,, & r\to\infty \\
~C_s + {C_v \over 3} r^3 \,, & r\to0.
\end{array} \ri.
\eea
Here $c_a,d_a$ and $K$ are finite $\o$-independent constants. The
$\o$-independent constants $C_s$ and $C_v$ can be related to $C_\o$ if we match the low-frequency behavior of equation \refeq{0T ir asympt} to the IR
behavior of equation \refeq{0T 0omega}. The outcome is:
\be\lab{CsCv}
C_s = \lim_{\o\to0} C_\o\, \o^{-a} \le( c_a - d_a K \o^{1-2a} \ri), \quad 
C_v = (1+2a) d_a \lim_{\o\to0} C_\o \, \o^{1+a}.
\ee
Imposing the boundary normalization of the zero-frequency wave function, taking its asymptotics from equation \refeq{0T 0omega}, at the cutoff
$r=r_Q$, implies
\be
1 = C_s + C_v r_Q^3 = \lim_{\o\to0} C_\o \o^{-a} \le[ c_a + d_a \le( {1+2a \over 3} r_Q^3 - K \ri) \o^{1+2a} \ri].
\ee
Hence, at low frequencies, the coefficient $C_\o$ evaluates to
\be\label{Comega}
C_\o \simeq \o^a \le[ c_a + d_a \le( {1+2a \over 3} r_Q^3 - K \ri) \o^{1+2a} \ri]^{-1} \simeq
{\o^{a} \over c_a}, \quad \o \to 0.
\ee

Substituting \refeq{Comega} into the equation for $\eta$, \refeq{kappa ir}, brings to the following result for the zero temperature diffusion constant:
\be\lab{0T kappa}
\eta \sim \lim_{\o\to0} \o^{2a} 
\ee

From (\ref{0T kappa}) we conclude that the zero temperature diffusion constant {\it vanishes for all backgrounds such that the metric in the IR satisfies}
\refeq{power}, with $a>0$. 
Such backgrounds include the conformal metric ($a=1$)
and all the scaling solutions ($a=\bar{a}$), for which the result (\ref{0T kappa}) agrees with the zero-frequency limit of the exact computations in Sections (\ref{0Tconf}) and (\ref{0Tscal}).

Thus, if we regulate the high-$\omega$ behavior by subtracting the zero-temperature
correlator, this does not affect the low-frequency limit.

Notice in particular that the small frequency limit of the correlator is completely determined by the IR part of the geometry, as one might have expected.

\subsection{Extension to confining backgrounds} \label{confiningsec}

We could perform the same analysis that brought to equations (\ref{Moo0})
and (\ref{0T kappa}) for backgrounds that, at $T=0$, are dual to confining gauge theories.
These are characterised by a string-frame scale factor that {\it grows} in the IR,
instead of decreasing to zero monotonically. The reason for this is that a necessary and sufficient condition for (holographic) confinement is that the scale factor has
a minimum at a non-zero value at some point $r_m$ in the bulk \cite{cobi}.
I
In 5D Einstein-dilaton theories with action (\ref{action2}), confinement requires a potential that
grows at least as (see \cite{ihqcd2} for details):
\be
V(\l) \simeq \l^{4/3} (\log \l)^{P} , \quad P>0, \qquad \l \to \infty.
\ee
Confining backgrounds for which the $r$-coordinate extends to infinity are characterized $0\leq P < 1$. For these backgrounds we have a power-law
behavior for the string frame scale factor in the IR\footnote{Note that the Einstein frame scale factor behaves exponentially in the same regime $\log
b_E\sim -r^{2a+1}$.}, like in equation (\ref{power}), of the form:
\be
b(r) \sim r^{-a}, \quad a = -{P\over 2-2P} \leq 0
\ee
The case of the Improved Holographic QCD model corresponds to $a=-1/2$.

The analysis of the correlators proceeds like in the previous sections. In particular, the high-frequency limit is still given by the universal
formula (\ref{Moo0}), since it is only sensitive to the UV. However, the analysis of the low-frequency limit for $a<0$ gives a result that at first
sight seems puzzling:
\be
 \eta \sim \lim_{\o\to0} \o^{|1+2a|-1} 
\ee
with a $\log^2\o$ correction in the special case $a=-1/2$.

The problem with this behavior is that, for the range $-1 < a< 0$ (which in fact contains the Improved Holographic QCD model), the diffusion constant
{\it diverges} at $T=0$. This suggests that the appropriate string solution is problematic in the IR.

A resolution of this puzzle lies in the fact that in such cases another embedding solution is relevant.
 This will be analyzed in a separate work \cite{langevin-confining}.

\section{The dressed Langevin evolution} \label{renormalized}

Having analyzed the finite- and zero-temperature Langevin correlators in the previous sections, we proceed to show that a natural choice for a UV-regulated propagator $\hat{G}_R(\o)$
that has the correct fall-off at large $\o$ is given, in the limit of $M_Q/\o \gg 1$,
by simply defining:

\be
\hat{G}_R(\o) = G_R(\o) - G_o(\o).
\ee

From the previous Sections we know that this procedure is consistent for all non-confining and scaling backgrounds, since it evidently takes care of the
leading divergent coefficient, and it does not modify the low-frequency behavior of the finite-$T$ correlators. At the end of this section we will also show that, after the subtraction, the dressed correlator
has the correct fall-off at large frequency to be consistent with a regular short-time evolution.

Here, by ``zero-temperature'' correlator, we mean a correlator evaluated in
the background whose metric coincides with the vacuum metric, but it is heated
up at a temperature $T$, i.e. the thermal gas background. This is necessary if we want to stay in the framework of the canonical ensemble, where $T$ is a fixed parameter. The thermal gas is a solution of Einstein equation that corresponds to a saddle point of the action, the true minimum being the black hole of the same temperature.

\subsection{Path integral treatment}

A physical correlated Langevin diffusion has force correlators that are regular functions of the time difference. In particular, as we have seen in Section \ref{vac}, an application of the formalism at $T=0$ (in $AdS$) would provide
a classical Langevin evolution (\ref{lange}) with a friction term analogous to the retarded correlator computed in $AdS$, that as we have seen is non-trivial and UV divergent.
However, we do not expect that a (free) heavy quark feels a force in the vacuum. This suggests that the proper physical coordinates of the heavy quark should be defined with respect to the $T=0$ background. We will do it explicitly below, using the path integral formalism.

We start from the Schwinger-Keldysh path integral, following the treatment found in \cite{iancu}
\be
Z=\int {\cal D}X_L {\cal D}X_R~e^{iS_R(X_R)-iS_L(X_L)},
\label{p1}\ee
with $X^{i}=X_0^i+\delta X^i$
where $X_0$ describes the classical motion of the dragging string and $\delta X^i$ is the fluctuation and
\be
S(X)=S(X_0)+\int d^2\xi P^{\a}_{i}\delta X^i-{1\over 2}\int d^2\xi ~G^{\a\b}_{ij}(X_0)\pa_{\a}\delta X^i\pa_{\b}\delta X^j+{\cal O}((\delta X)^3).
\label{p2}\ee
In this language, the trailing string equations are
\be
\pa_{\a}P^{\a}_i=0
\label{p3}\ee
and the classical equations for the fluctuations are
\be
\partial_{\a}\left(G^{\a\b}_{ij}(X_0)\pa_{\b}\delta X^j\right)=0.
\label{p4}\ee
It is convenient to Fourier transform
\be
\delta X^i(t,r)=\int {d\omega\over 2\pi} ~\zeta^{i}(\omega,r)~e^{-i\omega t}.
\label{p5}\ee
We can split the fields $\delta X_i$ into a boundary piece $x_i$ and the bulk piece $\delta X_i^b$. Integrating out the bulk pieces in (\ref{p1}) we obtain the boundary action as
\be
Z=\int {\cal D}X_L {\cal D}X_R~e^{iS(X_L,X_R)}=\int {\cal D}x_L {\cal D}x_R~e^{iS_{b}(x_L,x_R)},
\label{p6}\ee
with
\bea
iS_b&=&-i\int {d\omega\over 2\pi}\zeta^i_a(-\omega)G_{R}^{ij}(\omega)\zeta^j_r(\omega)
-{1\over 2}\int {d\omega\over 2\pi}\zeta^i_a(-\omega)G_{sym}^{ij}(\omega)\zeta^j_a(\omega)\nonumber\\
&&-i\int {d\omega\over 2\pi}\zeta^i_a(-\omega)P_i(\omega),
\label{p7}\eea
where we passed to the advanced and retarded basis, and $G_R$, $G_{sym}$ are the standard retarded and symmetric correlators, and $\zeta^i(\omega) \equiv \zeta^i(\omega, r_b)$ are the boundary values of the Fourier space variables.

We now use the identity
\bea
\exp\left[-{1\over 2}\int {d\omega\over 2\pi}\zeta^i_a(-\omega)G_{sym}^{ij}(\omega)\zeta^j_a(\omega)
\right]=det\left({G_{sym}^{-1}\over 2\pi}\right)^{1\over 2}\times
\label{p8}\eea
$$
\times
\int {\cal D}\xi^i~
\exp\left[-{1\over 2}\int {d\omega\over 2\pi}\xi^i(-\omega)(G^{-1}_{sym})^{ij}(\omega)\xi^j(\omega)+i\int {d\omega\over
2\pi}\xi^i(-\omega)\zeta_a^i(\omega)\right].
$$
The path integral becomes
\be
Z=\int d\mu~ e^{-S},
\label{p9}\ee
with
\be
d\mu=det\left({G_{sym}^{-1}\over 2\pi}\right)^{1\over 2}~{\cal D}\zeta^i_a {\cal D}\zeta^i_r {\cal D}\xi^i
\label{p10}\ee
and
\be
S=\int {d\omega\over 2\pi}\left[i\zeta^i_a(-\omega)\left(\xi^i(\omega)-P^i(\omega)-G_R^{ij}\zeta_r^j(\omega)\right)
-{1\over 2}\xi^i(-\omega)(G^{-1}_{sym})^{ij}(\omega)\xi^j(\omega)\right].
\label{p11}\ee

Integrating out $\zeta^i_a$ gives a functional $\delta$-function that imposes the Langevin equation, which can be written in configuration space as
\be
 \int_{-\infty}^{t}dt'G_{R}^{ij}(t,t')\zeta_r^j(t')+P^i(t)=\xi^i(t).
\label{p12} \ee
The noise variable $\xi^i$ has two-point function $G_{sym}^{ij}$.

The previous derivation holds also in the case of a quark propagating in the vacuum, since even in pure $AdS$ at zero temperature (and more generally in the non-conformal vacuum solutions) $G_{sym}$ and $G_R$ are non-trivial. In particular, in the vacuum solution we would get a Langevin equation like (\ref{p12}) with $G_R$ replaced by $G_R^o$, the vacuum retarded correlator,
and with noise distributed according go $G_{sym}^o$.

The fact that a particle in vacuum seems to undergo dissipation induced by quantum vacuum fluctuations seems rather unphysical. This is an indication that we need to modify the definition of the path integral in order to make sure that the average coordinate $\zeta_r(\omega)$ of a isolated particle in vacuum evolves according to its classical equation of motion, $P(\zeta_r(\omega)) = 0$. For simplicity we will assume that there is no external potential, thus the classical equation of motion is a linear term in the average coordinate: $P^i = {\cal M}^{ij}(\o) \zeta_{rj}(\omega)$ (e.g. for a non-relativistic particle $ {\cal M}^{ij}(\o) = M_Q \o^2 \delta^{ij}$ ).

We now proceed to define a path integral for a ``dressed'' string variable in such a way as to obtain the classical trajectory in the vacuum. We give this redefinition at the level of the boundary variables, i.e. after the bulk part of the string has been integrated out. The change in the path integral is composed of two parts:
\begin{enumerate}
\item
A change of variables:
\be\label{p13}
\zeta_r^i(\omega)=\left[\left({\cal M}(\omega) + G_{R}^o(\omega)\right)^{-1} {\cal M}(\omega)\right]^i_k \hat \zeta_r^k(\omega), \qquad
\zeta_a^i(\omega) = \hat\zeta_a^i(\omega).
\ee
We will take the dressed variables $\hat\zeta_r$ and $\hat\zeta_a$ as the true physical variables that describe the boundary quark.
\item A modified integration measure. We redefine the integration measure of the path integral to be used in (\ref{p6}) by:
\be\label{p14}
d\hat\mu = {\cal D}\zeta_a{\cal D}\zeta_r \exp\left[{1\over 2}\int {d\omega\over 2\pi}\zeta^i_a(-\omega)G_{sym}^{oij}(\omega)\zeta^j_a(\omega)
\right],
\ee
where ${\cal D}\zeta = \prod_\omega d^3\zeta^i(\omega)$ is the bare functional measure.
\end{enumerate}

We define the dressed boundary path integral at zero temperature as:
\be\label{p15}
{\cal Z}_o = \int d\hat\mu \exp^{iS^0_b(\zeta_r, \zeta_a)}
\ee
Using the explicit form of the action from (\ref{p7}) (with the zero-temperature $G_R$ and $G_{sym}$) we get immediately:
\be\label{p16}
{\cal Z}_0 = \int {\cal D}\hat\zeta_a {\cal D}\hat\zeta_r e^{i \int {d\omega\over 2\pi} \hat\zeta^i_a \, {\cal M}_i^k \zeta_{rk}} = {\cal N} \int
{\cal D}\hat\zeta_r \delta \left({\cal M}_i^k \, \hat \zeta^j_r (\omega)\right),
\ee
where ${\cal N}$ is a normalization constant. In other words, the vacuum path integral is concentrated on the classical trajectory with no dissipation for $\hat\zeta_r$, as required. This justifies our interpretation of the ``dressed'' variables $\hat\zeta$ as the physical ones describing the motion of the heavy quark.

Now we come to the non-trivial part, i.e. we define in the same way
the path integral at finite temperature:
\be\label{p17}
{\cal Z} = \int d\hat\mu \exp^{iS_b(\zeta_r, \zeta_a)}.
\ee
Performing the change of variables (\ref{p13}) and using the modified integration measure (\ref{p14}), it becomes:
\be\label{p18}
{\cal Z} = \int {\cal D}\hat\zeta_a {\cal D}\hat\zeta_r e^{-{1\over 2}\int {d\omega\over 2\pi} \hat \zeta_a^i \left[G_{sym} - G_{sym}^o\right]^{j}_i
\hat \zeta_{aj} + i \int {d\omega\over 2\pi} \hat \zeta_{ai} \left[\big({\cal M} + G_{R}\big)\big({\cal M} + G_{R}^o\big)^{-1}{\cal M}
\right]^i_{j}\hat \zeta_r^j}.
\ee
Following the same procedure that led to equation (\ref{p12}), we obtain a Langevin dynamics for the dressed variable $\hat\zeta_a$ defined by the
{\em dressed Langevin correlators}:
\be\label{p19}
\big(\hat G_R\big)^{i}_j = \left[\big({\cal M} + G_{R}\big)\big({\cal M} + G_{R}^o\big)^{-1}{\cal M} \right]^i_{j} - {\cal M}^i_j, \qquad \left(\hat
G_{sym}\right)^i_j = \left(G_{sym} - G^o_{sym}\right)^i_j.
\ee

For the symmetric correlator, the dressing simply amounts to the subtraction of the zero-temperature term. To obtain a simpler expression for the dressed $\hat G_R$, we first consider the non-relativistic case where
${\cal M}^i_j(\o) = M_Q \, \omega^2 \delta^i_j$. In this case, the first equation in (\ref{p19})
can be written as:
\be\label{p20}
\big(\hat G_R\big)^{i}_j = \left\{\left[\big(1 + G_{R}/M_Q\omega^2\big)\big(1 + G_{R}^o/M_Q\omega^2\big)^{-1} \right]^i_{j} - \delta^i_j\right\}M_Q\omega^2.
\ee
Throughout the paper we have been making the assumption that $\omega/M_Q \ll1$, in order for the holographic picture to be reliable. Since the zero temperature correlator behaves as $\omega^3$ both at large and small frequencies, we can expand the denominator and arrive at:
\be\label{p20b}
\left(\hat G_R\right)^{i}_j = \left(G_R - G^o_R\right)^i_j\, \left[1 + O\left(\omega/M_Q \right)\right].
\ee
Thus, the dressed retarded correlator also reduces, in this limit, to the subtracted correlator.

In the relativistic case, some extra work is required, since the effective mass matrix is not proportional to the identity, but is given by equation (\ref{mass}). Introducing the longitudinal and transverse projectors,
\be\label{p21}
\Pi^\perp_{ij} = \delta_{ij} - {v_i v_j\over v^2} , \qquad \Pi^\parl = {v_i v_j\over v^2}
\ee
we can write:
\be\label{p22}
M_{eff} = \gamma M_Q \Pi^\perp + \gamma^3 M_Q \Pi^\parl.
\ee
Correspondingly, we split the retarded correlator in a longitudinal and transverse part,
\be\label{22b}
G_R = \left(\Pi^\perp + \Pi^\parl \right) G_R \equiv G_R^\perp + G_R^\parl,
\ee
and similarly for $G^o_R$. Next, we rewrite the exrpression for $\hat G_R$ in (\ref{p19}) as:
\be\label{p23}
\hat G_R = (1 + {\cal M}^{-1}G_R ) (1 + {\cal M}^{-1} G_R^o)^{-1} {\cal M} - {\cal M}
\ee
and notice that ${\cal M}^{-1} G_R^o$ is small for small $\g \o M_Q$:
\be\label{p24}
{\cal M}^{-1} G_R^o = {G_R^{\parl o} \over \gamma^3 M_Q \o^2} + {G_R^{\perp o} \over \gamma M_Q \o^2} \sim \gamma \o M_Q \ll1.
\ee
Thus we can approximate $(1 + {\cal M}^{-1} G_R^o)^{-1} \simeq (1 - {\cal M}^{-1} G_R^o) $, and after some algebra equation (\ref{p23}) becomes:
\be\label{p25}
\hat G_R \simeq {\cal M}^{-1} \left(G_R - G^o_R\right) {\cal M} = (G_R - G^o_R)^\perp + (G_R - G^o_R)^\parl = (G_R - G^o_R).
\ee

Thus, with the assumption $\omega\ll M_Q$ we have the desired result that {\em the dressed correlators are simply obtained by subtraction of the zero temperature correlator both for the retarded and the symetric Green's functions:}
\be
\hat G_R \simeq G_R - G^o_R, \qquad \hat G_{sym} = G_{sym} - G^o_{sym}.
\ee

The correlators $\hat G_R$ and $\hat G_{sym}$ are the ones that should enter the physical Langevin diffusion in the plasma. $\hat G_R$ defines the physical spectral densities,
\be\label{}
\hat \rho (\o) = \rho(\o) - \rho^o(\o)
\ee

To complete the discussion, we establish the relation between $\hat G_R(\o)$ and $\hat G_{sym}(\o)$. For the bare correlator in the black hole background we have the Einstein relations at temperature $T=T_s$. It is reasonable to chose the temperature of the thermal gas background as the same $T_s$, such that two trailing string live in the same ensemble. Thus we have:
\be
G_{sym}(\o) = \coth\left({\o \over 2T_s}\right) {\rm Im}~G_R(\o) , \qquad G_{sym}^o(\o) = \coth\left({\o \over 2T_s}\right) {\rm Im}~G_R^o(\o)
\ee

If we make this choice of ensemble, it follows that the dressed correlators satisfies the standard Einstein relation:
\be
\hat G_{sym} = \coth\left({\o \over 2T_s}\right) {\rm Im}~\hat G_R.
\ee

 \subsection{UV behavior of the dressed spectral densities}

Having defined the physical Langevin spectral densities $\hat \rho (\o)$ at finite $T$ by subtracting the zero-temperature counterparts, our final task is to show that they display the correct large-$\o$ fall off, at least as $1/\o$.

As it is evident from Sections 2 and 3, $\hat \rho (\o) $ does not contain the $\o^3$ behavior at large frequency, since it cancels in the subtraction. However, this is not enough, as it could be that some terms that diverge linearly in $\o$ are left over after the subtraction, which would be unacceptable. This is not the case, as we show below.

Let us recall that the regime we are interested in is the high-frequency limit, $\gamma \o \gg 1/r_s$, and at the same time we need $\gamma \o r_Q \ll 1$ to be in the regime where the quark can be considered non-dynamical.

First, let us analyze the temperature-dependence of (\ref{Moo}). For this, we need the subleading behavior of the background functions entering the fluctuation equations near the boundary.
The crucial point is that temperature dependence manifests itself, in asymptotically $AdS$ backgrounds, as terms which all vanish as $r^4$ close to the boundary.

For asymptotically $AdS$ black holes, as $r\to 0$, we have
\be\label{s0}
f(r) \simeq 1 - {\cal C}r^4
\ee where ${\cal C}$ is a constant proportional to the product $ST$ of the black hole. Thus, from the expression (\ref{R}) for $R(r)$, we can write:
\be\label{s2}
R(r) \simeq {b^2(r)\over \gamma} \left[ 1 - {{\cal C}(T)\over 2} (\gamma^2 +1)r^4\right]
\ee
We also know, by a slight generalization of the discussion in \cite{gkmn2}, that close to the boundary\footnote{In general the scale factor will be temperature dependent. The only case when this is not so is that of $AdS$-Schwarzschild black holes}:
\be\label{s3}
b(r) = b_o(r) \left[1 + {\cal G}(T) g_1(r) r^4 \right],
\ee
where $ {\cal G}(T) $ is a temperature-dependent constant which controls the conformal anomaly, and $g_1(r)$ is a slowly-varying function that does not modify the power-law $r^4$ (otherwise consistency of Einstein's equations
would spoil the $AdS$ black hole asymptotics), but whose exact form depends on the specific model, and on whether the Einstein and string frame scale factor coincide.

The effective Schr\"odinger potential for the wave-function is given in equation (\ref{schr}). From the discussion above, it is clear that the only temperature-dependent corrections to the leading behavior, close to the boundary, scale as:
\be
\delta V(r) \sim r^4.
\ee
This would give rise to corrections to the wave functions that close to the boundary vanish as $r^4$, and when evaluated at $r_Q$ it can only give rise to
a temperature-dependence in the UV WKB coefficients which is of order $1/\o^4$ with respect to the leading behavior.

Close to the horizon, on the other hand, the expression (\ref{psihor}) does not get modified by finite $1/\o$ corrections. Thus, if we follow the matching
of the UV and horizon coefficients performed in Appendix \ref{wkbapp}, we conclude tht the only temperature dependence in (\ref{Moo}) enters at subleading $1/(\g\o)^4$, and there are no dangerous $1/(\g\o)^2$ terms.

We can then write, from the general expressions (\ref{Moo}) and (\ref{Moo0}) at large $\o$ (for e.g. the $\perp$ spectral density):
\be\label{s1}
\hat \rho^\perp(\o) \simeq { 1\over 2\pi^2 \ell_s^2} {\gamma^2 \o^3\over 1 + (\gamma \o r_Q)^2 }\Bigg[ \gamma r^2_{tp}(\o) R_{tp}(\o) - r^2_{tp,o}(\o)
b^2_o(r_{tp,o}(\o))\Bigg]\left[1 + {F(T,r_Q)\over (\g\o)^4}\right],
\ee
where $r_{tp,o}$ and $b_o(r)$ are respectively the WKB turning point and the scale factor at zero-temperature, the last brackets include the temperature-dependent corrections to (\ref{Moo}) discussed above, and we have dropped the $(\gamma \o \r_Q)^4$ terms in the denominator\footnote{these terms can easily be evaluated explicitly from the WKB calculation in Appendix \ref{wkbapp}. For example, the full expression in the case of IHQCD can be found in \cite{langevin-1}. They depend on h(r) evaluated at the turning point, and by the argument presented below they cannot contribute more than $1/\o$ in the UV to the subtracted spectral densities.}.

The term in the first bracket also gives rise to terms that grow no faster than $1/\o^4$: for large frequency the turning point is close to the boundary, $r_{tp} \sim 1/(\gamma \o)\ll r_s$, both at zero and at finite temperature; the functional dependence on $r_{tp}$ scales as $r_{tp}^4$ due
to equations (\ref{s0}-\ref{s3}), and the turning point is determined by the Schr\"odinger potential $ V(r)$, which is also modified at $\cO (r^4)$. Thus the turning points at a {\em fixed} frequency $\gamma \o$ are shifted at most by the same fractional amount,
\be\label{s4}
r_{tp} = r_{tp,o}\left[1 + g_2(r_{tp},T) r_{tp}^4\right],
\ee
where again $g_2(r,T)$ is another slowly varying function
which can be determined from the Schr\"odinger potential.

Collecting results (\ref{s2}-\ref{s4}) in the expression (\ref{s1}) we arrive at:
\be\label{s5}
\hat \rho^\perp \simeq { 1\over 2\pi^2 \ell_s^2} {1\over 1 + (\gamma \o r_Q)^2 } {g(\o;T,r_Q) \over \gamma^2 \o} ,
\ee
where $g(\o;T,r_Q)$ is a temperature-dependent function, slowly-varying (with respect to a power law) in $\o$, which collects all the various contributions discussed above. This is the final result, that shows that the dressed correlator we have defined has a well-defined Fourier transform, thus it satisfies all the appropriate dispersion relations and provides a consistent Langevin process at short times.

\section{Examples of dressed Langevin correlators}

In this Section we present the numerical evaluation of the full dressed correlators, as a function of $\o$, in two interesting models: the conformal
case, and the non-confining scaling backgrounds.

\subsection{Dressed correlators in the conformal case}
\FIGURE[b]{
\includegraphics[width=6.7cm]{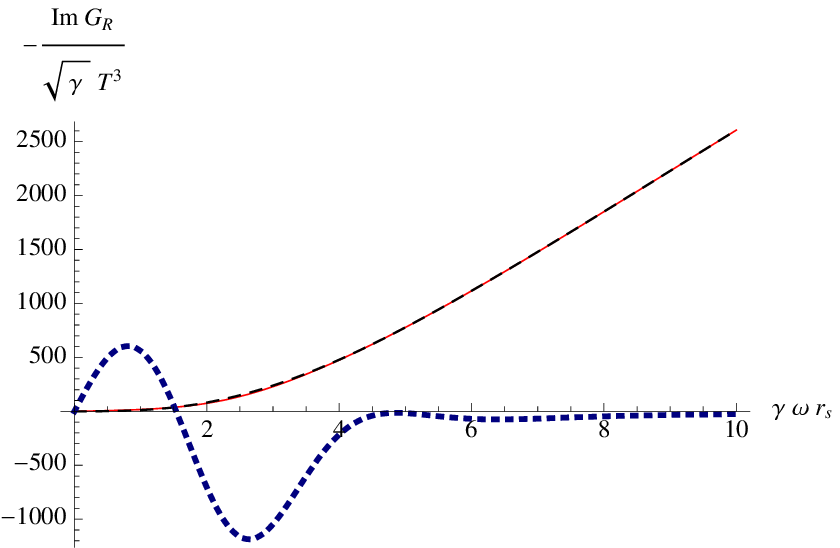} \hspace{0.7cm}
\includegraphics[width=6.5cm]{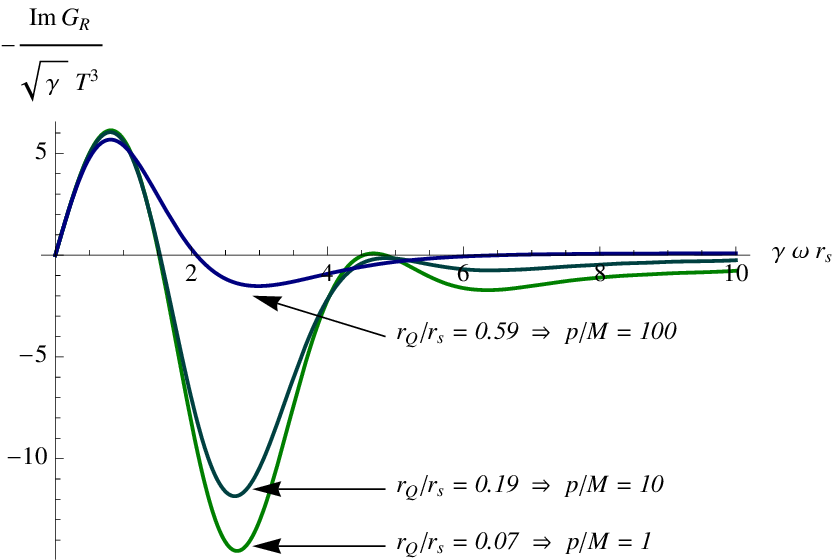}\\\\
\hspace{2.5cm}(a) \hspace{7cm}(b)
\caption{The imaginary part of the retarded correlator, rescaled by $\sqrt\g T^3$, is shown as a function of the dimension-less frequency $\g \o r_s$.
For both figures we set $r_h/r_Q=17$, implying $M_Q/T=20$. In (a) we plot the numeric result for the finite temperature correlator (red plain curve),
the zero temperature analytic result (black dashed curve) and the renormalized correlator magnified by a factor 100 (blue dotted curve), for
$r_Q/r_s=0.19$, meaning $p/M_Q=10$. In (b) we consider three different values for the ratio $p/M_Q$, $p/M_Q=1,10,100$ ($r_Q/r_s=0.07,0.19,0.59$), and show
the renormalized correlator for each of them.}\lab{fig GR conf}}
In the conformal case all temperatures above zero are equivalent. However, since we introduce a cutoff determined by the quark mass as in
\refeq{mass-cutoff}, the mass-to-temperature ratio is related to the ratio $r_h/r_Q = \sqrt\g r_s/r_Q$ (and by the velocity, through $\g$) as follows:
\bea
{M_Q \over T} = { \sqrt{\l_{{\cal N}=4}} \over 2} {r_h \over r_Q}.
\eea

In Figure \ref{fig GR conf} we show the imaginary part of the retarded correlator, rescaled by $\sqrt\g T^3$, and as a function of the dimension-less
frequency $\td \o=\g \o r_s$, for a specific value of $M_Q/T$ and different values of $p/M_Q=\sqrt{\g^2-1}$. On the left, the renormalized correlator is
compared to the finite temperature and zero temperature correlators for fixed momentum, while on the right we show the renormalized correlators for
different momenta.

We chose to use the dimension-less frequency $\td \o$, because it encodes the whole dependence of the rescaled correlator on the temperature and on
the velocity, in the limit of infinitely massive quarks (when the cutoff is sent to zero). This follows from the fact that the fluctuation equations
only depends on the frequency, temperature, velocity and radial coordinate through this combination and through the ratio dimension-less coordinate
$x=r/r_s$ (as described in Subsection \ref{conformal diffusion}).

However, in the case of a finite cutoff, there appears an additional dependence on the velocity, through the ratio $r_Q/r_s=x_Q$, leading to different
curves for different momenta. This stems from the fact that the flux must now be evaluated at $r=r_Q$, rather that at $r=0$ as in the infinite mass
case, where the dependence on the velocity and temperature is all taken into account by the rescaled correlator and frequency. Hence, as we vary the
velocity in the finite mass case, we change the value of the momentum-to-mass ratio $p/M_Q$, which is related to the ratio $r_Q/r_s$ by
\bea
{p \over M_Q} = \sqrt{ {16 \over \l_{{\cal N}=4}} \le( {M_Q \over T} \ri)^4 \le( {r_Q \over r_s} \ri)^4 - 1 }.
\eea
Moreover, we note that the cutoff $r_Q$ should be kept fixed if we want to keep the quark mass fixed.

In Figure \ref{fig GR conf}(a) we note that the imaginary part of the retarded correlator displays the expected behavior at high frequencies: cubic at
intermediate frequencies (namely for frequencies still much lower than $1/\g \o r_Q$) and linear at higher frequencies. The renormalized correlator
quickly goes to zero as the frequency grows.

The dressed correlators plotted in Figure \ref{fig GR conf}(b) show the dependence on the velocity that arises due to the finite mass. The
rescaled dressed correlators depend both on the velocity and on the temperature only through the ratio $r_Q/r_s$.

\subsection{Dressed correlators in scaling backgrounds}
\FIGURE[t]{
\includegraphics[width=6.7cm]{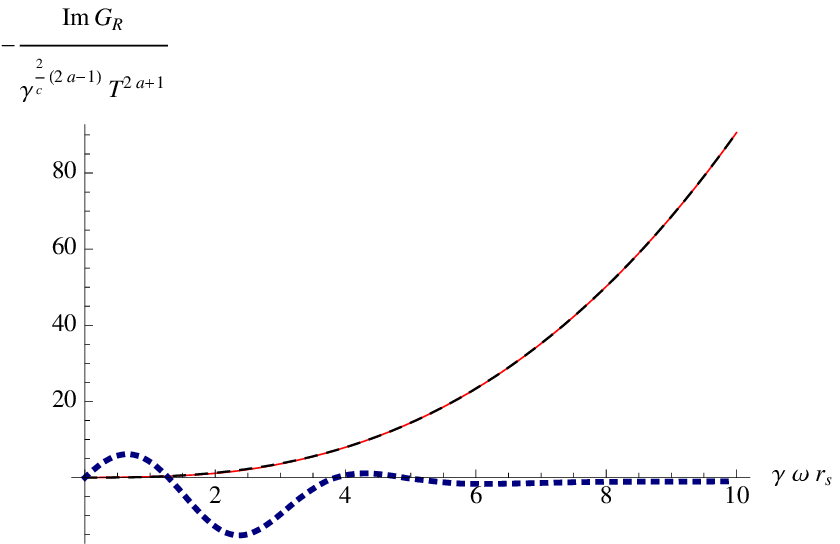} \hspace{0.7cm}
\includegraphics[width=6.5cm]{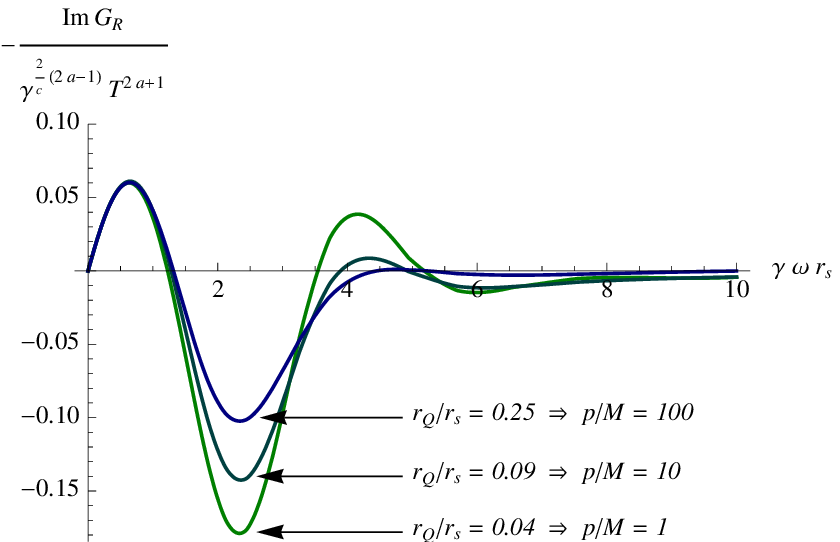}\\\\
\hspace{2.5cm}(a) \hspace{7cm}(b)
\caption{The imaginary part of the retarded correlator, rescaled by $\g^{2(2\bar a-1)/c}T^{2\bar a+1}$, is shown as a function of the dimension-less frequency
$\g \o r_s$. For both figures we chose the background parameters such that $\bar a=5/6$ and $c=23/5$ in $p+1=5$ dimensions (this means $k=0$ and
$\d^2=1/9$). We also set $r_h/r_Q=30$, implying $\ell M_Q/(\ell T)^{2\bar a-1}=2$. As for the conformal case, in (a) we plot the numeric result for the
finite temperature correlator (red plain curve), the zero temperature analytic result (black dashed curve) and the renormalized correlator magnified
by a factor 100 (blue dotted curve), for $r_Q/r_s=0.09$, meaning $p/M_Q=10$. In (b) we consider three different values for the ratio $p/M_Q$,
$p/M_Q=1,10,100$ ($r_Q/r_s=0.04,0.09,0.25$), and show the renormalized correlator for each of them.}\lab{fig GR scal}}

The scaling backgrounds described by the metric functions in \refeq{scbk}, with $\bar a>0$, share a similar behavior with the conformal $AdS_5$ model.
In this Subsection we show the numerical results for the dressed correlator in this class of models. The zero temperature correlator is derived
analytically, using the exact expression \refeq{rho scaling exact}, while the finite temperature correlator is computed numerically.

The presence of a finite mass introduces a scale in these models. The mass-to-temperature ratio is given in terms of the quantity $r_h/r_Q$ by
\bea
{\ell M_Q \over (\ell T)^{2\bar a-1}} = {2(2\bar a-1) \over c} {\ell^2 \over \ell_s^2} \le({ r_h \over r_Q }\ri)^{2\bar a-1}.
\eea

We restrict to the region of parameters of the scaling backgrounds where $\bar a>1/2$, in order to produce a consistent spectral density, as discussed
in Section \ref{0Tscal}. The plots shown in figure \ref{fig GR scal} represent the imaginary part of the retarded correlator divided by a temperature
and velocity dependent factor, $\g^{2(2\bar a-1)/c}T^{2\bar a+1}$, which encodes the whole dependence on $T$ and $v$ in the infinite mass limit. As in
the conformal case, there is an extra dependence on $v$ (at fixed $T$ and $M_Q$) when the mass is finite. This dependence arises through the ratio
$r_Q/r_s$, appearing in the $\cO\le( (\g \o r_Q)^2 \ri)$ corrections -- and higher orders -- in \refeq{rho scaling} and \refeq{rho scaling exact}.

The momentum-to-mass ratio can be expressed in terms of $r_Q/r_s$ as
\bea
{p \over M_Q} = \sqrt{ \le( {2 \pi \over 2\bar a-1} {\ell_s^2 \over \ell^2} {\ell M_Q \over (\ell T)^{2\bar a-1}} \ri)^{c \over 2\bar a-1} \le( {c \over 4 \pi} {r_Q \over r_s} \ri)^c -1 }.
\eea
So at fixed ${\ell M_Q / (\ell T)^{2 \bar a-1}}$ (or, equivalently, at fixed $r_h/r_Q$) the spectral densities corresponding to different momenta
$p/M_Q$
(or different $r_Q/r_s$) will have a different behavior. This is shown explicitly in figure \ref{fig GR scal}.

\vskip 3cm 
\addcontentsline{toc}{section}{Acknowledgements}
\section{Acknowledgements}\label{ACKNOWL}

We would like to thank Yukinao Akamatsu, Jorge Casalderrey-Solana and Edmond Iancu for useful conversations. This work was partially supported by a European Union grant FP7-REGPOT-2008-1-CreteHEP Cosmo-228644, and PERG07-GA-2010-268246. The work of LM was supported in part by MICINN and FEDER under grant FPA2008-01838 and by the Spanish Consolider-Ingenio 2010 Programme CPAN (CSD2007-00042).

\newpage
 \addcontentsline{toc}{section}{Appendices}
\appendix
\section*{Appendix}
\section{The high-frequency expansion} \label{wkbapp}
In this appendix we review and expand the result obtained in \cite{langevin-1} about the high-frequency behavior of the spectral densities. We will perform the detailed analysis only for transverse modes. The same method can be obtained for longitudinal modes with minor modifications. The analysis we provide here is general, and covers both asymptotically $AdS$ solutions, and scaling backgrounds.

The starting point is the wave equations (\ref{H19}):
\be\label{Psieq}
\de_r\Big(\, R \,\de_r \Psi(r,\o)\Big) + {\o^2 b^4 \over R} \Psi(r,\o) =0
\ee
where
\be
R(r) = \sqrt{(f(r)-v^2)(b^4(r) f(r) - C^2)}, \qquad C = v b^2(r_s).
\ee
This equation can be put in Schroedinger form with the redefinition:
\be
\psi= R^{1/2}(r) \Psi,
\ee
in terms of which it reads:
\be \label{schr}
-\psi''(r) + V(r) \psi(r) =0, \qquad V(r) =
-\frac{\o^2 b^4}{R^2} + \half \big(\log R\big)'' + \frac14
\big(\log R \big)^{'2}.
\ee
The first term in the potential is the only important one for large $\o$, except in a small region close to the boundary, where $R \sim b^2/\g$: since $b(r)$ diverges as a power-law for small $r$, close to the boundary the $\o$ independent terms grow approximately as $r^2$, whereas the first term stays constant. Thus, for large $\o$, we can neglect the $\o$-independent terms everywhere except in the region $0< r \lesssim 1/(\g \o)$. Elsewhere, one can use the WKB approximation to compute the solution ({\it WKB region}). On the other hand, it is possible to approximately solve equation (\ref{schr}) analytically for $0< r \ll r_s$, ({\it UV region}), which for large frequency $\g \o \gg 1/r_s $ overlaps with the WKB region. This allows to determine the coefficient $\Psi_h$ of equation (\ref{psihor}), thus the spectral density (\ref{IMGR}). This procedure is explained in detail in the next Subsection.
\subsection{The WKB wavefunctions}
In the range $0 < r < r_s$ we can identify three regions:
\begin{itemize}
\item {\it WKB region}: $1/(\g \o) \ll r \leq r_s$
\item {\it UV region}: $r_Q< r \ll r_s$.
\item {\it Overlap region}: $ 1/(\g\o) \ll r \ll r_s$.
\end{itemize}
The overlap region exists for large frequencies, $\g\o \gg 1/r_s$.
Notice that the WKB region contains the horizon. The UV boundary $r=r_Q$ is finite for finite quark mass, whereas $r_Q \to 0$ for strictly non-dynamical quarks.

Below we solve for $\Psi(r,\o)$ separately in the UV and WKB regions, and match the coefficients in the overlap region.
\subsubsection*{WKB region}
In this region, we can approximate
\be \label{Vwkb}
V(r) \simeq {\o^2 b^4\over R^2}
\ee
In the WKB approximation, the independent solutions of equation (\ref{schr}) are
$$\psi = V^{-1/4}\,\exp \left(\pm i \int \sqrt{V(r)}\right)$$.
Using the explicit form (\ref{Vwkb}) for the potential, we can write $\Psi = R^{-1/2}\psi$ as:
\be\label{Psiwkb}
\Psi_{WKB} = \Psi_h {b(r_s) \over b(r)} \exp i \o \int^r dr' {b^2(r') \over R(r')}
\ee
where the normalization has been chosen for later convenience. The choice of a purely ingoing wave is the one that gives the correct behavior at the horizon: as $r\to r_s$, $R(r)$ behaves as:
\be
R(r) \sim 4\pi T_s\, b^2(r_s)\, (r_s-r), \qquad r\to r_s
\ee
thus we find, close to the horizon, the appropriate in-falling solution:
\be
\Psi_{WKB} \simeq \Psi_h (r_s - r)^{-i\o/4\pi T_s}.
\ee
\subsubsection*{UV region}
For small $r \ll r_s$ the effect of temperature is negligible, as $f\simeq 1$. In this region we can approximate
\be\label{Ruv}
R \simeq b^2/\g,
\ee
 and equation (\ref{schr}) becomes simply:
\be\label{scrhUV}
-\psi'' + {\sqrt{R}''\over \sqrt{R}}\psi = \o^2 \gamma^2 \psi.
\ee
To find approximate solutions of this equation, we distinguish two situations:
\begin{enumerate}
\item {\it power-like case: $b(r)\sim (\ell/r)^p$}\\
 In case $b(r)$ is a pure power law at small $r$, $b(r) \simeq (\ell/r)^p$ (as is the case in asymptotically $AdS$ backgrounds, or in the scaling backgrounds analysed in Section \ref{scaling},
then $(\sqrt{R})''/\sqrt{R} \simeq p(p+1)/r^2$ and the solution for $\Psi = \sqrt{R} \psi$ is simply given in terms of Bessel functions:
\be\label{PsiUV}
\Psi_{UV}(r) = {A \over \sqrt{R(r)}} \sqrt{r}\left[J_{p+1/2}(\g\o r) + i N_{p+1/2}(\g\o r)\right]
\ee
This linear combination chosen in such a way that it connects with the ingoing wave (\ref{Psiwkb}) in the overlap region: for $ r/r_s \ll 1 $ but $\g \o r \gg 1$ we can use the large-argument asymptotic of the Bessel functions:
\be
J_\nu(x) \simeq \sqrt{2\over \pi x}\sin\left( x - \nu\pi/2 - \pi/4\right) , \qquad N_\nu(x) \simeq \sqrt{2\over \pi x}\cos\left( x - \nu\pi/2 - \pi/4\right)
\ee
Using this approximation as well as equation (\ref{Ruv}) to compare the two forms (\ref{PsiUV}) and (\ref{Psiwkb}) in the overlap region, we find the relation between the coeffifients $A$ and $\Psi_h$ (up to an oveall constant phase):
\be\label{match}
A \sqrt{2\over \pi \o} = \Psi_h b(r_s)
\ee
On the other hand $A$ is fixed by requiring unit normalization $\Psi(r_Q)=1$. For small $\gamma \o r_Q$ we can use the small argument expansion of the Bessel functions:
\be\label{besssmall}
J_\nu(x) \simeq {x^\nu \over 2^\nu \Gamma(\nu +1)}, \qquad N_\nu(x) \simeq {2^\nu \Gamma(\nu) \over \pi x^\nu}
\ee
to obtain the normalization (using $R\simeq b^2/\g \simeq (\ell/r)^{2p}/\gamma$):
\be\label{unit}
1 = A \,{\sqrt{\g} \over \ell^p} {i 2^{p+1/2} \Gamma(p+1/2) \over \pi (\g \o)^{p+1/2}}\left[ 1 + O\left((\g\o r_Q)^2\right)\right]
\ee
From equations (\ref{match}-\ref{unit}) we determine $\Psi_h$, which inserted in equation (\ref{IMGR}) determines the spectral density:
\be
\rho^\perp \simeq {\ell^{2p} \over 2\pi \ell_s^2} {\g^{2p}\over 2^{2p+1} \Gamma^2(p+1/2)} \o^{2p+1} \left[ 1 + O\left((\g\o r_Q)^2\right)\right].
\ee

The conformal case of $AdS$ asymptotics corresponds to $p=1$. For this value of $p$ we recover the results found in \cite{langevin-1}. In this case the Bessel functions have a simple expression in terms of sines and cosines, and it is not hard to specify the first correction to the leading result:
\be\label{rhoconf}
\rho^\perp_{conf} \simeq {\ell^{2} \over \pi \ell_s^2} {\gamma^2 \o^3 \over 1 + (\g\o r_Q)^2 + O\left((\g\o r_Q)^4\right)}.
\ee
 \item {\it corrected power-law $b(r)$}\\
Let us now take $p=1$, but consider the more general case of the asymptotics (\ref{relaxed}), where $h(r)$ does {\it not} approach a constant as $r\to 0$ (otherwise we would fall back in the previous case with $p=1$). In the following we will restrict to approximate $AdS$ asymptotics with $p=1$, but the method can be easily generalized for any $p$.

With the asymptotics (\ref{relaxed}), the Schroedinger potential in the UV becomes:
\be\label{Vrelaxed}
V \simeq {2\over r^2}\left(1 - {2 r h' \over h} + r^2 {h'' \over h}\right) - \g^2\o^2
\ee
If the $h$-dependent terms in the parenthesys are small for small $r$, we may think that we can still approximate the solution as a linear combination of the form:
\be\label{psiwrong}
\psi(r) = A r^{1/2}\left[J_{3/2}(\g\o r) + i N_{3/2}(\g\o r)\right]
\ee
Expanding this for $r\to 0$ leads to $\psi \simeq C_s /r + C_v r^2 $
for some constants $C_s$ and $C_v$, which would lead to $\Psi \sim R^{-1/2}(r) ( C_v/r + C_v r^2) \sim h^{-1/2}(r) ( C_v + C_v r^3) $. This is not, however, the correct behavior for small $r$, which can be obtained directly by integrating equation (\ref{Psieq}) for $r\ll 1/(\g\o)$ neglecting the $\o^2$ term:
\be\label{smallr}
\Psi \sim C_s + C_v \int^r_0 {dr'\over R(r')} \qquad r\to 0.
\ee
In order to get a consistent solution in the whole UV region, we must correct equation (\ref{psiwrong}) as follows \cite{langevin-1}:
\be \label{psiright}
\psi(r) \simeq r^{1/2}\left[A_2 F_2(r,\o) J_{3/2}(\g\o r) + i A_1 F_1(r,\o) N_{3/2}(\g\o r)\right]
\ee
where $F_1$ and $F_2$ are functions that are slowly varying, and behave as:
\be\label{Fas}
F_1 \sim (r^2 R(r))^{1/2}, \quad F_2 \sim (r^2 R(r))^{-1/2}, \qquad r\to 0.
\ee
With these ansatz, the wavefunction $\Psi = R^{-1/2}\psi$ has the correct behavior at small $r$, as one can see by expanding the Bessel functions and comparing with equation (\ref{smallr}).

On the other hand, as the solution enters the overlap region ($r > 1/(\g\o)$) , the functions $F_1$ and $F_2$ must approach two constants, in order to match the WKB solution.
 Since they are slowly varying, we can estimate these constants to be the values
\be\label{Fas2}
F_1(r) \to \bar{F}_1 \simeq (r^2 R)^{1/2}\Big|_{r_{tp}} , \qquad F_2 \to \bar{F}_2 \simeq (r^2 R)^{-1/2}\Big|_{r_{tp}}, \qquad r \gg 1/(\g\o),
\ee
 where $r_{tp}$ is the classical turning point of the Scrhoedinger equation, and it marks the ``boundary'' between the UV region and the WKB region. For large $\o$ it is approximately $r_{tp}\simeq \sqrt{2}/(\gamma\o)$ as can be seen from equation (\ref{Vrelaxed})
Now the matching in the overlap region leads to:
\be\label{match2}
A_1 \bar{F}_1 = A_2 \bar{F}_2 = \Psi_h b(r_s)\sqrt{{\pi \o\over 2}}
\ee
and the unit normalization $Psi(r_Q)=1$ is:
\be
1 =i A_1 {F_1(r_Q)\over R^{1/2}(r_Q)} r^{3/2}_Q N_{3/2}(\g \o r_Q) \left[1 - i {A_2\over A_1} \left({F_2 J_{3/2}\over F_1 N_{3/2}}\right)\Bigg|_{rQ}\right]
\ee
To lowest order in $r_Q$, using the asymptotics (\ref{Fas}) and those of the Bessel functions, this gives:
\be
A_1 = \sqrt{\pi/2}(\g\o)^{3/2}\left[ 1 + (\g\o r)^2 + O\left((\g\o r_Q)^4\right)\right]^{-1}.
\ee
From equation (\ref{Fas2}) and (\ref{match2}) we can determine $\Psi_h$, which inserted in equation (\ref{IMGR}) gives:
\be\label{rhoconf2}
\rho^\perp \simeq {\ell^{2} \over \pi \ell_s^2} \gamma^3 \o^3 \,r_{tp}^2 R(r_{tp}) \left[ 1 + (\g\o r_Q)^2 + O\left((\g\o r_Q)^4\right)\right]^{-1}.
\ee
A similar procedure gives $\rho^\parl(\o)$, the only difference being an extra factor $\gamma^2$ in the numerator.

\end{enumerate}

 \subsection{WKB at zero temperature} \label{wkbapp0}

At zero temperature the blackness function $f(r)\equiv 1$ and there is no horizon. Correspondingly, we have
\be
R(r) = {b^2(r) \over \gamma}.
\ee
 We consider backgrounds in which the conformal coordinate reaches $+\infty$ in the interior, where the scale factor behaves as a power-law,
\be
b(r) \sim r^{-a}, \quad a>0, \qquad r \to \infty
\ee
The only change with respect to the previous Section is that
the WKB region now covers the deep IR, and the WKB wavefunction (\ref{Psiwkb}) is given by
\be
\Psi_{WKB}^o = C_\o r^{a} e^{i\gamma \o}
\ee
One then repeats exactly the same steps as in the previous Section in the UV, to determine $C_\o$, and the result is the same as (\ref{rhoconf2}), with $R(r)$ replaced by $b^2(r)/\gamma$.

\section{The boundary effective action}
In this appendix we compute the boundary effective action for the world-sheet fluctuations, at the boundary point $r_Q$. We concentrate on the real part of the quadratic on shell action, which should reproduce the corresponding action for the fluctuations of a relativistic particle moving at constant speed in four dimensions. We will see that this is the case only in the limit $\g\o r_Q \to 0$.

We restrict our attention to (possibly corrected) $AdS$ asymptotics. Also, we only consider here one transverse fluctuation; generalizaztion to longitudinal ones is straightforward. We start with the quadratic action for world-sheet fluctuations, equation (\ref{H16}).
\be \label{H16app}
S^{(2)}=-{1\over 4\pi \ell_s^2}\int_{-\infty}^{+\infty} d\tau \int_{r_Q}^{r_s} dr \left[ H^{\a\b} \partial_{\a} \Psi\partial_{\b} \Psi\right].
\ee
Evaluating the action on shell should produce boundary terms that, in the limit $r_Q \to 0$, should match the form of boundary diff-invariant counterterms for the source term in the fluctuation, i.e. $\Psi(r_Q, t)$. The only such term, in a homogeneous medium, is the one coming from kinetic action for a relativistic particle,
\be\label{B2}
S_{bdr} = M_Q \int d\tau \sqrt{\dot{X}^\mu \dot{X}^\nu h_{\mu\nu}},
\ee
where $h_{\mu\nu}$ is the induced boundary metric. Expanding (\ref{B2}) around a constant speed trajectory with a transverse fluctuation, $X^0 = \tau$, $\vec{X}= \xi_0 + \vec{v} t + \Psi(t) \vec{u}^\perp$ with $\vec{u^\perp} \cdot \vec{v}=0$, we obtain (going to Fourier space):
\be\label{S2bdr}
S^{(2)}_{bdr} = {M_Q\over 2\g}\int_{-\infty}^{+\infty} {d\o\over 2\pi} \gamma^2 \o^2 |\Psi(\o)|^2
\ee

Now we evaluate equation the bulk action, (\ref{H16app}), on the in-falling solution $\Psi(r,\o)$, but relaxing the unit boundary normalization, i.e. we take $\Psi(r_Q,\o) = \Psi(\o)$ to be arbitrary. After an integration by parts and use of the field equations for $\Psi(r,\o)$, as well as $H^{rr} = R(r)$, we obtain:
\be\label{osac}
S^{(2)} = {1\over 4\pi \ell_s^2} \left(\int {d\o\over 2\pi} \Psi^*(\o) R(r_Q) \de_r \Psi(\o) - \int {d\o\over 2\pi} \Psi^*(r_s,\o) R(r_s) \de_r \Psi(r_s,\o)\right)
 \ee
The second term arises from the world-sheet horizon, and can be regarded as an IR modification to the quark mass. We are interested here in the boundary contribution, so we will focus on the first term only.

In order to evaluate the first term, we use the high-frequency solutions found in Appendix \ref{wkbapp}. We need to separate the fluctuations in two sets, according to the value of $\o$.
\begin{enumerate}
\item $\o \gg 1/(\g r_Q)$\\
In this case $r_Q$ belongs to the WKB region, and we can use equation (\ref{Psiwkb}), where now $\Psi_h b(r_s) \to \Psi(\o)$.
Inserting this expression in the first term in equation (\ref{osac}) we find:
\be\label{notmass}
S_{UV}^{(\o > 1/(\g r_Q))} \simeq {1\over 4\pi \ell_s^2} \int_{|\o| > 1/(\g r_Q)} {d\o\over 2\pi} \left[\left({b'\over b}{R\over b^2}\right)_{r=r_Q} + i \o \right] |\Psi(\o)|^2
\ee
Comparing with equation (\ref{S2bdr}) we see that the real part of this expression {\em does not correspond to a boundary kinetic term} for the string endpoint as it is of order zero in $\o$. Rather, it gives rise to a harmonic potential.

\item $\o \ll 1/(\g r_Q)$\\
In this case, $r_Q$ is in the UV region, but not in the boundary region. We can use $\Psi(r) = R^{-1/2}\psi(r)$, with $\psi(r)$ given by (\ref{psiright}),
where now the coefficients $A_1,A_2$ are arbitrary functions of $\o$. It is convenient to rewrite the wavefunction in the form:
\be\label{Psiuvreg}
\Psi = \Psi(\o) \, \sqrt{r/r_Q \over R(r)/R(r_Q)} {F_1(r) \over F_1(r_Q)} {N_{3/2}(\g\o r)\over N_{3/2}(\g\o r_Q)} \left[1 + i {\bar{F}_2 (\o) \over \bar{F}_1 (\o)} {F_2(r) \over F_1(r)} { J_{3/2}(\g\o r) \over N_{3/2}(\g \o r)} \right],
\ee
Since we are in the region $r_Q \ll 1/(\g \o)$, we can use the approximation (\ref{Fas}) for the functions $F_1$ and $F_2$:
\be
 \sqrt{r/r_Q \over R(r)/R(r_Q)} {F_1(r) \over F_1(r_Q)} \simeq \left(r /r_Q\right)^{3/2}.
\ee
 Moreover, we can use the small argument expansion $N_{3/2}(x) \simeq -\sqrt{2/\pi} x^{-3/2} -x^{1/2}/ \sqrt{2\pi} + \ldots$ both in the numerator and denominator (it is necessary to go to first subleading order to obtain a non-zero result):
\be
{N_{3/2}(\g\o r)\over N_{3/2}(\g\o r_Q)} \simeq (r_Q / r)^{3/2}\left[1 + {\g^2 \o^2 r^2\over 2} + O\left(\g \o r\right)^4\right]
\ee
The term inside the parenthesys of equation (\ref{Psiuvreg}) , on the other hand, is approximated by:
\be
1 - i {r^2 R(r) \over r_{tp}^2 R(r_{tp})} \left[ {\g^3 \o^3 r^3\over 3} + O\left( (\g\o r)^5 \right)\right]
\ee
as $r\to 0$, as it results from the expasion (\ref{besssmall}) and from eqs. (\ref{Fas}-\ref{Fas2}).
Putting everything together we obtain, in the limit of small $\g \o r_Q$:
\be
\Psi (r,\o) \simeq \Psi(\o) \left[1 + {\g^2 \o^2 r^2\over 2} + O\left(\g \o r\right)^4\right] \left[ 1 + iO\left( (\g\o r)^3 \right) \right].
\ee
The real part of the UV boundary action in (\ref{osac}) is therefore:
\be
S_{UV}^{(\o < 1/(\g r_Q))} \simeq {1\over 4\pi \ell_s^2} r_Q R(r_Q) \int_{|\o| < 1/(\g r_Q)} {d\o\over 2\pi} \, \g^2 \o^2 |\Psi(\o)|^2
\ee
This has the same form of a mass term, with the effective quark mass given by:
\be
M_Q = {1\over 2\pi \ell_s^2} r_Q R(r_Q) \g \simeq {1\over 2\pi \ell_s^2} r_Q b^2(r_Q) \g
\ee
Notice that this diverges as we remove the cutoff, $r_Q \to 0$: as it should be, the divergence is cancelled precisely by the boundary counterterm (\ref{B2}), as was noted in \cite{langevin-1}.
Also, notice that in the conformal case we recover the relation $M_Q = (\ell^2/2\pi \ell_s^2) 1/r_Q$.
\end{enumerate}

\section{The neutral scaling solutions} \label{scalingapp}

We start with the action
\be
S=M^{p-1}\int d^{p+1}x\sqrt{-g}\left[R-{1\over 2}(\partial\phi)^2+V(\phi)\right]
\label{1}\ee
with
\be
V=2\Lambda e^{-\delta\phi}
\label{2}\ee

In the domain-wall coordinate system,
\be
 \ud s^2 = e^{2A}\left[-f(r)\ud t^2 +\ud x_i\ud x^i \right] + \frac{\ud r^2}{f(r)}\,,
 \label{3}\ee
we have the scaling solutions

 \be
e^A=r^{2\over (p-1)\d^2}\,,\qquad e^{\delta\phi}={\Lambda\d^4\over \left({2p\over p-1}-\d^2\right)}~r^{2}\,,\qquad f=1-\left({r_0\over r}\right)^{{2 p\over (p-1)\d^2}-1}\,,
\label{4}\ee

The metric can be rewritten by a charge of coordinates
\be
w=r^{1-{2\over (p-1)\d^2}}\sp t\to {t\over \Big |1-{2\over (p-1)\d^2}\Big|}\sp x^i\to {x^i\over \Big |1-{2\over (p-1)\d^2}\Big|}
 \label{5}\ee
 as conformal to an $AdS$-like black hole,
 \be
 ds^2=e^{2\chi}~{{dw^2\over f}-fdt^2+dx^idx^i\over w^2}
 \label{6} \ee
with
\be
 e^{2\chi}=2{{2p\over p-1}-\d^2\over \left({2\over p-1}-\d^2\right)^2}{1\over V(\phi)}\sp f=1-\left({w\over w_0}\right)^{2p-(p-1)\d^2 \over 2-(p-1)\d^2}
\label{7}\ee

From this we can rewrite the metric in the conformal (and Einstein) frame
\be
ds^2=b(w)^2\left[-fdt^2+dx^idx^i+{dw^2\over f}\right]
\label{8}\ee
with
\be
b=\left({\ell\over w}\right)^a\sp e^{\phi}=\Lambda_0\left({w\over \ell}\right)^d\sp f=1-\left({w\over w_0}\right)^c\sp w_0={c\over 4\pi T}
\label{9}\ee
\be
a={2-(p-1)\d^2\over 2}\sp c={2p-(p-1)\d^2 \over 2-(p-1)\d^2}\sp d={2(p-1)\d\over (p-1)\d^2-2}
\label{10}\ee

For $\d=0$ we obtain the $AdS$$_{p+1}$ solution.
Note that always $2p-(p-1)\d^2>0$ in order to satisfy the Gubser bound. Therefore, for $\d>0$, the sign of all $a,c,d$ is always the same.
Also, in the regime $\d^2<{2\over p-1}$, $c\in [p,\infty)$, while in the regime $\d^2>{2\over p-1}$, $c\in (-\infty,0]$

We now consider a dragging string moving in this metric. In the following we can set $\Lambda_0=1$ as this does not affect the results.
 There are several options here , because we need to pass from the Einstein to the string metric and this depends on whether $\phi$ is the string dilaton or not.
 We parametrize this dependence by
 \be
 b_s=b_E~e^{k\sqrt{1\over 2(p-1)}\phi}
 \label{11}\ee
with $k=1$ corresponding to the string dilaton and $k=0$ to a completely different scalar.

We obtain
\be
b_s=\left({\ell\over w}\right)^{\bar a}\sp \bar a=a-{kd\over \sqrt{2(p-1)}}
\label{12}\ee

The turning point $w_s$ of the dragging solution is given by
\be
f(w_s)=v^2~~~\to~~~w_s=w_0(1-v^2)^{1\over c}={c\over 4\pi T}(1-v^2)^{1\over c}
\label{13}\ee
From now we use the string frame scale factor $b_s$ that we still call $b$.

We can compute
\be
{b'(w_s)\over b(w_s)}=-{\bar a\over w_s}=4\pi T{\bar a\over c}(1-v^2)^{-{1\over c}}\sp {f'(w_s)\over f(w_s)}=-{4\pi T\over v^2}(1-v^2)^{c-1\over c}
\label{14}\ee
and
\be
4\pi T_s=\sqrt{f(w_s)f'(w_s)\left(4{b'(w_s)\over b(w_s)}+{f'(w_s)\over f(w_s)}\right)}=4\pi T(1-v^2)^{c-2\over 2c}\sqrt{1-\left(1-{4\bar a\over c}\right)v^2}
\label{15}\ee
Note that as $\bar a c>0$, the last square root never vanishes.

The appropriate coupling to the dilaton is the one that makes $4\bar a=c$, and then
\be
k={\sqrt{2(p-1)}\over 4}~{2(p-1)^2\d^4- 7(p-1)\d^2 - 2 (p-4)\over 2-(p-1)\d^2}
\label{16}\ee

We now move to study the equations for the two point functions
\be
\partial_w\left[ \sqrt{(f-v^2)(b^4f-C^2)}\,\,\partial_w\left(\delta X^{\perp}\right)\right]+{\omega^2b^4\over \sqrt{(f-v^2)(b^4f-C^2)}}\,\delta X^{\perp}=0
\label{17}\ee
\be
\partial_w\left[{1\over Z^2}\sqrt{(f-v^2)(b^4f-C^2)}\,\,\partial_w\left(\delta X^{\parallel}\right)\right]+{\omega^2b^4\over Z^2\sqrt{(f-v^2)(b^4f-C^2)}}\delta X^{\parallel}=0
\label{18}\ee
with
\be
Z\equiv b^2\sqrt{f-v^2\over b^4f-C^2}\sp C=vb(w_s)^2
\label{19}\ee

Define
\be
x={w\over w_s}={w\over w_0}(1-v^2)^{-{1\over c}}\sp \partial_w={1\over w_s}\pa_x\sp f(w)-v^2=(1-v^2)(1-x^c)
\label{20}\ee
\be
b^4(w)f-C^2=\left({\ell\over w_s}\right)^{4\bar a}~{1-(1-v^2)x^c-v^2 x^{4\bar a}\over x^{4\bar a}}
\label{21}\ee
\be
Z=\sqrt{1-v^2}\sqrt{1-x^c\over 1-(1-v^2)x^c-v^2 x^{4\bar a}}
\label{22}\ee
We then have
\bea
x^{2\bar a}\partial_x\left[x^{-2\bar a}\sqrt{(1-x^c)(1-(1-v^2)x^c-v^2x^{4\bar a})}\pa_x\left(\delta X^{\perp}\right)\right]+&&\nonumber\\
+{\Omega^2 \over \sqrt{(1-x^c)(1-(1-v^2)x^c-v^2x^{4\bar a})}}\delta X^{\perp}&=&0
\label{23}\eea
with
\be
\Omega={w_s~\omega\over \sqrt{1-v^2}}={c(1-v^2)^{2-c\over 2c}\over 4\pi}~{\omega\over T}
\label{24}\ee
For $\d=0$, $p=4$, $k=0$ the dependence on $v$ cancels and it reduces to the $AdS$$_5$ case.

When $4\bar a=c$, $Z^2=1-v^2$ and the equation simplifies to
\be
x^{c\over 2}\partial_x\left[x^{-{c\over 2}}(1-x^c)\pa_x\left(\delta X^{\perp}\right)\right]
+{\Omega^2 \over (1-x^c)}\left(\delta X^{\perp}\right)=0
\label{25}\ee

\end{document}